\documentclass[journal]{IEEEtran}

\usepackage{ucs}
\usepackage[utf8x]{inputenc}
\usepackage[cmex10]{amsmath}
\usepackage{cite, amsfonts, amssymb, amsthm, bm, bbm, graphicx, relsize, multirow, booktabs, color, colortbl,soul}

\usepackage[american]{babel}
\usepackage[T1]{fontenc}
\usepackage{algorithmic, algorithm}
\usepackage{tracklang}
\usepackage{glossaries}
\usepackage{upgreek}

\theoremstyle{remark}

\DeclareMathOperator{\rank}{rank}

\DeclareMathOperator{\diag}{diag}
\DeclareMathOperator{\trace}{Tr}

\newcommand{\test}{\mathrel{\underset{H_0}{\overset{H_1}{\gtrless}}}}

\newcommand{\e}{\mathrm{e}}
\renewcommand{\i}{\mathrm{i}}

\newcommand*{\herm}{{\mathsf{H}}}
\newcommand*{\transp}{{\mathsf{T}}}
\newcommand{\cc}{\cellcolor[gray]{0.9}}
\graphicspath{{figures/}}

\title{Foundations of MIMO Radar Detection Aided by Reconfigurable Intelligent Surfaces}
\author{ Stefano~Buzzi,~\IEEEmembership{Senior Member,~IEEE}, Emanuele~Grossi,~\IEEEmembership{Senior~Member,~IEEE}, Marco~Lops,~\IEEEmembership{Fellow,~IEEE}, Luca~Venturino,~\IEEEmembership{Senior Member,~IEEE}  
\thanks{The work of S. Buzzi, E. Grossi and L. Venturino was supported by the research program ``Dipartimenti di Eccellenza 2018--2022'' sponsored by the Italian Ministry of Education, University, and Research (MIUR).}
\thanks{S.~Buzzi, E.~Grossi and L.~Venturino are with the Department of Electrical and Information Engineering (DIEI), University of Cassino and Southern Lazio,  03043 Cassino, Italy, and with Consorzio Nazionale Interuniversitario per le Telecomunicazioni, 43124 Parma, Italy (e-mail: buzzi@unicas.it; e.grossi@unicas.it; l.venturino@unicas.it).}
\thanks{M.~Lops is with the Department of Electrical and Information Technology (DIETI), University of Naples Federico II, 80138 Naples, Italy, and with Consorzio Nazionale Interuniversitario per le Telecomunicazioni, 43124 Parma, Italy (e-mail: lops@unina.it).}
\thanks{Part of this work has been submitted for possible presentation at the 2022 IEEE Radar Conference and the 2022 International Conference on Acoustics, Speech, \& Signal Processing.}
}

\newacronym{cbap}{CBAP}{contention-based access period}
\newacronym{cphy}{CPHY}{control physical}
\newacronym{dmg}{DMG}{directional multi-gigabit}
\newacronym{dti}{DTI}{data transmission interval}
\newacronym{mcss}{MCSs}{modulation and coding schemes}
\newacronym{scphy}{SCPHY}{single-carrier physical}
\newacronym{qo}{QO}{quasi omni-directional}

\newacronym{ssp}{SSP}{scheduled service period}
\newacronym{stf}{STF}{short training field}
\newacronym{cef}{CEF}{channel estimation field}
\newacronym{br}{BR}{beam refinement}
\newacronym{iss}{ISS}{initiator sector sweep}
\newacronym{rss}{RSS}{responder sector sweep}
\newacronym{sls}{SLS}{sector level sweep}
\newacronym{ssw-fb}{SSF}{sector sweep feedback}
\newacronym{ssw-ack}{SSA}{sector sweep acknowledgement}

\newacronym{ap}{AP}{access point}
\newacronym{cdma}{CDMA}{code-division multiple-access}
\newacronym{ds/cdma}{DS/CDMA}{direct-sequence code-division multiple-access}
\newacronym{ofdma}{OFDMA}{orthogonal frequency-division multiple-access}
\newacronym{pam}{PAM}{pulse amplutude modulation}
\newacronym{bpsk}{BPSK}{binary phase shift keying}
\newacronym{mac}{MAC}{multiple-access channel}
\newacronym{bc}{BC}{broadcast channel}
\newacronym{bs}{BS}{base station}

\newacronym{far}{FAR}{false alarm rate}
\newacronym{rcs}{RCS}{radar cross-section}
\newacronym{cfar}{CFAR}{constant false alarm rate}
\newacronym{lfm}{LFM}{linear frequency modulated}
\newacronym{mvdr}{MVDR}{minimum variance distortionless response}
\newacronym{pri}{PRI}{pulse repetition interval}
\newacronym{amf}{AMF}{adaptive matched filter}
\newacronym{mf}{MF}{matched filter}
\newacronym{mf-pd}{MF-PD}{matched-filter peak detector}
\newacronym{wmf}{WMF}{whitening matched filter}

\newacronym{csi}{CSI}{channel state information}
\newacronym{los}{LOS}{line-of-sight}
\newacronym{nlos}{NLOS}{non-line-of-sight}
\newacronym{mimo}{MIMO}{multiple-input multiple-output}
\newacronym{siso}{SISO}{single-input single-output}

\newacronym{ris}{RIS}{reconfigurable intelligent surface}

\newacronym{dfrc}{DFRC}{dual-function radar-communication}
\newacronym{papr}{PAPR}{peak-to-average power ratio}
\newacronym{pn}{PN}{pseudo-noise}
\newacronym{psd}{PSD}{power spectral density}

\newacronym{iaa}{IAA}{iterative adaptive approach}
\newacronym{iaa-apes}{IAA-APES}{iterative adaptive approach for amplitude and phase estimation}
\newacronym{iic}{IIC}{iterative interference cancellation}

\newacronym{sic}{SIC}{successive interference cancellation}
\newacronym{snr}{SNR}{signal-to-noise ratio}
\newacronym{sinr}{SINR}{signal-to-interference-plus-noise ratio}
\newacronym{inr}{INR}{interference to noise ratio}
\newacronym{zf}{ZF}{zero-forcing}
\newacronym{mmse}{MMSE}{minimum mean square error}
\newacronym{nrmse}{NRMSE}{normalized root mean square error}

\newacronym{iid}{i.i.d.}{independent and identically distributed}
\newacronym{awgn}{AWGN}{additive white Gaussian noise}

\newacronym{lr}{LR}{likelihood ratio}
\newacronym{llr}{LLR}{log-LR}
\newacronym{ml}{ML}{maximum likelihood}

\newacronym{lhs}{LHS}{left-hand side}
\newacronym{rhs}{RHS}{right-hand side}
\newacronym{lrs}{LRS}{local reference system}

\newacronym{aic}{AIC}{Akaike information criterion}
\newacronym{bic}{BIC}{Bayesian information criterion}
\newacronym{gic}{GIC}{generalized information criterion}
\newacronym{glrt}{GLRT}{generalized likelihood ratio test}
\newacronym{crb}{CRB}{Cram\'er-Rao bound}
\newacronym{cdf}{CDF}{cumulative distribution function}
\newacronym{pdf}{PDF}{probability density function}
\newacronym{pmf}{pmf}{probability mass function}

\begin{document}
\bstctlcite{BSTcontrol}
\maketitle

\begin{abstract}
A reconfigurable intelligent surface (RIS) is a nearly-passive flat layer made of inexpensive elements that can add a tunable phase shift to the impinging electromagnetic wave and are controlled by a low-power electronic circuit.  This paper considers the fundamental problem of target detection in a RIS-aided multiple-input multiple-output (MIMO) radar. At first, a general signal model is introduced, which includes the possibility of using up to two RISs (one close to the radar transmitter and one close to the radar receiver) and subsumes both a monostatic and a bistatic radar configuration with or without a line-of-sight view of the prospective target. Upon resorting to a generalized likelihood ratio test (GLRT), the design of the phase shifts introduced by the RIS elements  is formulated as the maximization of the probability of detection in the location under inspection for a fixed probability of false alarm, and suitable optimization algorithms are proposed. The performance analysis shows the benefits granted by the presence of the RISs and shed light on the interplay among the key system parameters, such as the radar-RIS distance, the RIS size, and location of the prospective target. A major finding is that the RISs should be better deployed in the near-field of the radar arrays at both the transmit and the receive side. The paper is  concluded by discussing some open problems and foreseen applications. 
\end{abstract}

\begin{IEEEkeywords}
MIMO Radar, LOS/NLOS, Bi/Mono-Static, Target Detection, Reconfigurable Intelligent Surfaces.
\end{IEEEkeywords}

\section{Introduction}
\Glspl{ris} are attracting a huge interest from researchers, as they allow to realize smart radio environments~\cite{di2020smart,Geoffrey-Ye-2020,liu2021reconfigurable}. An \gls{ris} is a nearly-passive low-cost planar structure made of engineered materials with tunable electromagnetic characteristics; current implementations include reflectarrays, trasmitarrays, liquid crystal surfaces, and software-defined meta-surfaces~\cite{foo2017liquid,liaskos2018using,hum2013reconfigurable}. Such device does not emit any power of its own and only aims to manipulate existing waves to alter the wireless propagation channel. An \gls{ris} differs from a specular reflector or a scattering object, as its elements can be individually controlled  by a low-power external logic to redirect the incident electromagnetic wave towards an arbitrary (anomalous) direction or a specific location. This focusing  mechanism resembles that of a phased array and can enhance the signal reception at a desired destination. Differently from amplify-and-forward relaying, the signal boost is obtained without using a radio-frequency chain or consuming power for amplification\footnote{Active \glspl{ris} able to amplify and redirect the incident electromagnetic signal  have been  studied in~\cite{Larsson-2021,ActiveRIS2021} to compensate for the two-hop attenuation in the \gls{ris}-aided link; however, these surfaces are not considered here.} or adding a processing delay/noise or requiring a half-duplex operation~\cite{RISvsRELAY-2020}. 
In light of their low hardware footprint, the \glspl{ris} can be easily deployed in the environment; for example, they can be integrated into the facades and the roof of a building, the walls and the ceiling of a room, and the case of a laptop.  \gls{ris}-empowered radio environments can be  exploited in all kind of wireless services, including communication, localization, and radar operations, opening up new opportunities. As it is meaningful to create an RIS-aided indirect link only if the cascade of the source-RIS and RIS-destination channels is sufficiently good, the system engineer must preliminary verify a good location for the RIS placement.

In the past years, the \glspl{ris} have been mainly investigated in wireless communications to enhance the network performance. In~\cite{Rui-2019}, a \gls{ris} assists the communications between a multiantenna \gls{bs} and multiple single-antenna users, and  the total
transmit power is minimized via joint active and passive beamforming, subject to a \gls{sinr} constraint for each user. In~\cite{howmany}, the phase shifts of the \gls{ris} can only take value in a finite set, and the relation between the set cardinality and the rate degradation is investigated. In \cite{beyond2020}, the phase shifts of the \gls{ris} are employed in conjunction with the communication transmitter to encode a message, showing that this scheme achieves a larger capacity than that obtained when the \gls{ris} is designed only to maximize the received \gls{snr}. The work in~\cite{joint2020Ye}, instead, considers a point-to-point \gls{mimo} link and performs an alternating optimization  of the transmit precoder and the phase shifts of the \gls{ris} to minimize the symbol error rate. The channel estimation in a \gls{ris}-aided link has been tackled in~\cite{mirza2021channel}, while other studies have assessed the \gls{ris} benefits in the context of wireless power transfer~\cite{zhao2020wireless}, 
antenna design~\cite{2021-Tulino}, and coverage extension~\cite{zeng2020reconfigurable}. The reader may refer to~\cite{di2020smart, Geoffrey-Ye-2020, liu2021reconfigurable} for a more comprehensive overview.  

The \glspl{ris} can also be exploited for user localization in mobile networks~\cite{2020-Wymeersch-Localization-and-Mapping}. In~\cite{Alouini-Localization}, a scenario with one \gls{bs}, one mobile device, and one \gls{ris} is considered, and the achievable localization and orientation performance with synchronous and asynchronous signalling schemes are studied. The use of an \gls{ris} may also allow to perform joint synchronization and localization  without using two-way transmission, even if the mobile user is equipped with one antenna~\cite{fascista2020ris}. The problem of joint localization and communication is instead considered in~\cite{RIS_loc_and_comm}, which presents an \gls{ris} design based on the use of hierarchical codebooks and feedback from the mobile station. 
  
More recently, researchers have started investigating the \gls{ris} benefits in  \gls{dfrc} systems. 
In~\cite{2020-Wang-Spectrum-Sharing}, the  radar  detection  probability is maximized by optimizing  the   transmit  beamformer of the \gls{bs} and the phase shifts of the \gls{ris},  subject to \gls{sinr}  and  power  constraints. Instead, the transmit waveforms of the \gls{bs} and  the phase shifts of the \gls{ris} are chosen in~\cite{joint_waveform} to mitigate the multiuser interference under a beampattern constraint. 

With regard to the fundamental problem of radar target detection, in our earlier contribution~\cite{Grossi2021ris} we assumed that the radar is capable of forming (a) two transmit  beams pointing towards the prospective target and the \gls{ris} and one receive beam pointing towards the prospective target only or, viceversa, (b) one transmit beam pointing towards the prospective target only and two receive beams pointing towards the prospective target and the \gls{ris}. Under such a scenario, a theoretical  analysis  is  carried  out  for  closely-  and widely-spaced  (with  respect  to  the  location of the prospective target)  radar  and  \gls{ris} deployments,  showing that  large  gains  can  be  provided  by the nearby \gls{ris}, and initial hints on the optimal \gls{ris} placement are provided. In~\cite{aubry2021reconfigurable}, the  case where no direct path between the radar and the prospective target exists is considered, mimicking the companion situation where the \gls{ris} provides an indirect link between a communication source and a destination that would be otherwise not reachable. Finally, \cite{MIMO_sensors} represents a first contribution to the scenario considered here and, although passing over a number of relevant effects and situations, as detailed in the sequel of this contribution, makes the basic point that an \gls{ris} can indeed aid the \gls{mimo} radar detection. 

\begin{figure*}[t]	
	\centering	
	\centerline{\includegraphics[width=\textwidth]{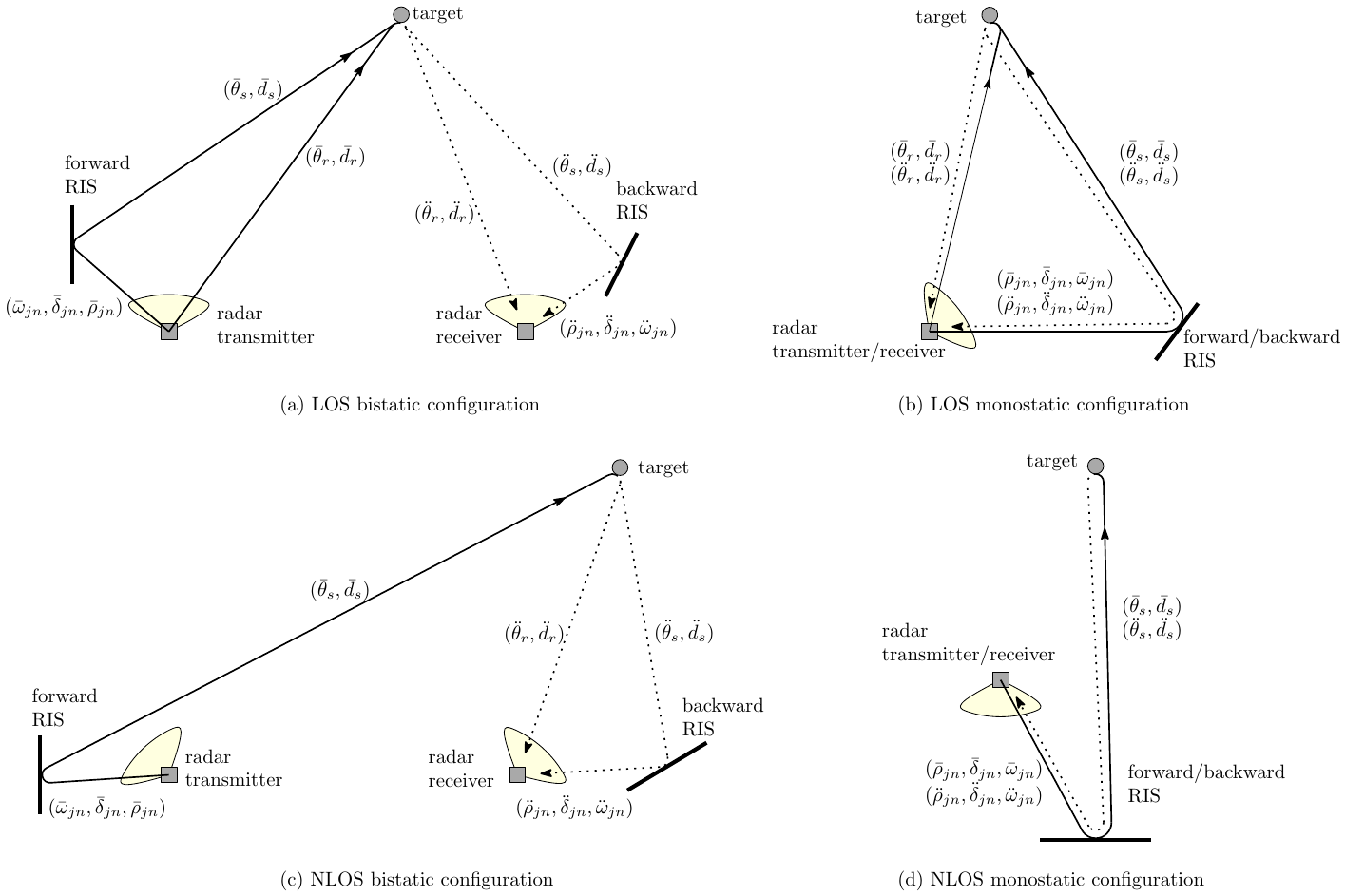}}	
	\vspace{-0.2cm}\caption{Examples of bistatic and monostatic system geometries. In configurations (a) and (b), the radar transmitter and receiver have a LOS view of the prospective target; the indirect paths granted by forward/backward RISs can be exploited here to improve the performance that the radar would have alone. In configuration (c), the radar transmitter does not have a LOS view of the prospective target; the forward RIS grants here a NLOS path for its illumination. In configuration (d), the radar only relies on the forward/backward RISs to illuminate the prospective target and capture its echo.} \label{fig:system}	
\end{figure*}

\subsection{Contribution and paper structure}
The bottom line of the current studies is that suitably deployed \glspl{ris} can modify the wireless channel response, thus providing novel degrees of freedom to the system design. In this work, we focus on the classical target detection problem~\cite{KayBook_vol2,book-Richards,Grossi-2016} and investigate  under which conditions the \glspl{ris}  may boost the performance of an \gls{mimo} radar with closely-spaced antennas, while preserving its key functions, i.e., illuminating a large angular sector and focusing on a desired point by only relying on receive signal processing.  For a given \gls{ris} size, the distance covered by its re-directed wave is tied to the power of the incident wave, which in turn depends on the power emitted by the active source and the length of the source-RIS hop. Accordingly, several usage contexts are possible: for example, radars for air-traffic, coastal, and marine surveillance could be aided by nearby \glspl{ris} deployed on the same structure hosting the radar, a nearby building/wall, or the ground itself, while wireless access points with integrated radar and communication capabilities can use the same \glspl{ris} to support both functions.

With reference to the scenarios outlined in Fig.~\ref{fig:system}, we first derive  a novel signal model, wherein the \gls{mimo} radar transmits a set of orthogonal waveforms and, assisted by a \emph{forward} and/or a \emph{backward} \gls{ris} operating as a reflecting mirror, verifies the presence/absence of a prospective target in a given location under inspection. The  model includes monostatic,  bistatic, \gls{los}, and  \gls{nlos} radar configurations and accounts for the presence of up to four paths from the transmitter to the prospective target to the receiver. Also, we distinguish among the two relevant situations that the forward and backward \glspl{ris} are in the near- or far-field of the radar transmitter and receiver, respectively.  

Next, we derive the \gls{glrt} receiver with respect to the unknown target response and tackle the design of the phase shifts introduced by the reflecting elements by maximizing the target detection probability for a fixed probability of false alarm: interestingly, the optimization problem turns out to be {\em separable}, i.e., the forward and backward \glspl{ris} can be designed independently. Also, if the forward (backward) \gls{ris} and radar transmitter (receiver) are in each other's far-field the optimum phase shifts are found in closed form; instead, in a near-field situation, the optimization problem is strongly NP-hard in general and can be approximately solved through an alternate maximization or by resorting to a convex relaxation. Interestingly, this latter approach provides a randomized solution whose expected value is at least $\pi/4\approx 0.785$ times the optimum. The special case in which the radar is monostatic and the same \gls{ris} provides both a forward and a backward indirect path is also examined, and an ad-hoc algorithm to select the phase shift of each reflecting element is proposed. 

A performance analysis has been carried out to assess the impact of the radar-\gls{ris} distance, the \gls{ris} size, and the location under inspection.  The results show that only a marginal gain (as compared to the case where the radar operates alone) can be obtained if radar and \gls{ris} are in each other's far-field; this confirms the intuition in~\cite{Grossi2021ris} that an \gls{ris} should be better placed in proximity of the radar transmitter or receiver. Conversely, a large performance gain can be obtained under a near-field radar-\gls{ris} deployment; interestingly, such gain almost doubles when the prospective target is in the view of both the forward and backward \glspl{ris}, as compared to the case where only a single surface illuminates/observes it. The performance of a monostatic radar with a single \gls{ris} (the simplest and perhaps most realistic and cost-effective configuration) is also contrasted to that of an {\em ideal} system where the phase shifts of the reflecting elements are changed in between the transmission and reception phases, showing that these configurations are substantially equivalent. 

The paper is organized as follows. In Sec.~\ref{SEC:System-description},  the system description  is presented. In Sec.~\ref{SEC:System-design}, the design of the phase shifts introduced by the \gls{ris} elements is studied. In Sec.~\ref{Sec:analysis}, some examples are given to assess the performance of an \gls{ris}-aided radar. Finally, concluding remarks and hints for future developments are provided in Sec.~\ref{Sec:conclusions}.
 
\subsection{Notation} 
Column vectors and matrices are denoted by lowercase and uppercase boldface letters, respectively. The symbols $(\cdot)^{*}$, $(\cdot)^{\transp}$ and $(\cdot)^{\herm}$ denote conjugate, transpose, and conjugate-transpose, respectively. $\bm{I}_{M}$ is a $M\times M$ identity matrix, while  $A_{ij}$ and $a_{i}$ denote the $(i,j)$-th entry of the matrix $\bm{A}$ and the $i$-th entry of the vector $\bm{a}$, respectively. $\bm{A}\succeq 0$ means that $\bm{A}$ is Hermitian positive semidefinite. $\trace\{\bm{A}\}$ and $\|\bm{a}\|$ denote the trace of the square matrix $\bm{A}$ and the Frobenius norm of the vector $\bm{a}$, respectively. $\rank\{\bm{A}\}$ is the rank of the matrix $\bm{A}$. 
$\diag\{\bm{a}\}$ is the $M\times M$ matrix containing the entries of $M$-dimensional vector $\bm{a}$ on the main diagonal and zero elsewhere. The symbol $\otimes$ denotes the Kronecker product, ${\rm E}[\,\cdot\,]$ the statistical expectation, and $\i$ the imaginary unit. 

\section{System description}\label{SEC:System-description}
We consider an \gls{ris}-aided \gls{mimo} radar aimed to detect a prospective target in a given location under inspection. The  radar transmitter and receiver are equipped with $\bar{N}_{r}\geq1$ and $\ddot{N}_{r}\geq1$ closely-spaced elements, respectively, arranged into a planar array;\footnote{Henceforth, the diacritical marks $\bar{(\cdot)}$ and $\ddot{(\cdot)}$ are used to distinguish between the transmit and receive sides, respectively: this notation mimics the solid and dotted line-styles adopted in  Fig.~\ref{fig:system}.} the building element of each array can be a single antenna or a subarray module composed itself of multiple antennas. The radar emits $\bar{N}_{r}$ orthogonal and equal-power waveforms, one from each radiating  element. A planar \gls{ris} (referred to as the forward \gls{ris}) can help the radar transmitter illuminate the prospective target;  this surface is composed of $\bar{N}_{s}\geq 1$ elements, also called unit cells or meta atoms, that can change the phase of the incident electromagnetic field while reflecting it. Similarly, a planar \gls{ris} (referred to as the backward \gls{ris}) with $\ddot{N}_{s}\geq 1$ tunable reflecting elements can help the radar receiver capture the power scattered by a target.\footnote{The following signal model and design methodology still remain valid if each \gls{ris} element retransmits (rather than reflects) a phase-shifted version of the incident electromagnetic field~\cite{2021-Tulino}.} A data link is assumed to exist between the radar control unit, where the system design is carried out, and each RIS.   

The \glspl{ris} can modify the response of the environment to the waveforms emitted by the radar; in particular, a prospective target can be illuminated by the radar transmitter (direct path) and/or by the forward \gls{ris} (indirect path); likewise, the reflections originated  by the target may reach the radar receiver through a direct path and/or a indirect path bouncing on the backward \gls{ris}. As shown in Fig.~\ref{fig:system}, up to four echoes may be observed, corresponding to a direct illumination and a direct reflection,  a indirect illumination and a direct reflection, a direct illumination and a indirect reflection, and  a indirect illumination and a indirect reflection. he model considered in this work (see Section~\ref{Sec:rx-signal}) encompasses both a bistatic and a monostatic radar; in this latter case, the forward and backward \glspl{ris} may collapse into a single surface redirecting the incident waves from the radar transmitter and the target, as shown in Figs.~\ref{fig:system}(b)-(d). Also, the model encompasses both a \gls{los} and a \gls{nlos} radar; in the former case, both the transmitter and the receiver already have a direct view of the prospective target and the additional \gls{ris}-assisted paths can be exploited to improve the performance that the radar would have alone; in the latter case, the indirect paths granted by the \glspl{ris} may extend the field of view of the radar, as shown in Figs.~\ref{fig:system}(c)-(d). We underline that the configuration in Fig.~\ref{fig:system}(d) may be of interest also to  replace an expensive radar transceiver that would be required to cover the region of interest with a low-cost feeder that sends/receives signals via a re-configurable surface capable of electronically-tunable beamforming~\cite{Grossi2021ris}.

\subsection{Geometric parameters}
The transmit/receive/reflecting arrays are located on the $(y,z)$-plane of a Cartesian local reference system, with elements oriented towards the positive $x$-axis. 
At the transmit side, we define the following quantities:\footnote{We follow the standard notation that the azimuth angle of a point $\bm{p}\in\mathbb{R}^3$ is the angle between the $x$-axis and the orthogonal projection of the vector pointing towards $\bm{p}$ onto the $(x,y)$-plane, which takes values in $[-\pi,\pi)$ and is positive when going from the $x$-axis towards the $y$-axis; also, the elevation angle is the angle between the  vector pointing towards $\bm{p}$ and its orthogonal projection onto the $(x,y)$-plane, which takes values in $[-\pi/2,\pi/2)$ and is positive when going towards the positive $z$-axis from the $(x,y)$-plane.}
\begin{itemize}
	\item $\bar{d}_{r}$, $\bar{\theta}_{r}^{\rm az}$, and $\bar{\theta}_{r}^{\rm el}$ are the range and the azimuth and elevation angles of the target from the reference element of the transmit array of the radar, respectively;
	
	\item $\bar{d}_{s}$, $\bar{\theta}_{s}^{\rm az}$, and $\bar{\theta}_{s}^{\rm el}$ are the range and the azimuth and elevation angles of the target from the reference element of the forward \gls{ris}, respectively;
	
	\item $\bar{\delta}$, $\bar{\rho}^{\rm az}$, $\bar{\rho}^{\rm el}$, $\bar{\omega}^{\rm az}$, $\bar{\omega}^{\rm el}$  are the length, the azimuth and elevation angles of departure, and the azimuth and elevation angles of arrival, respectively, of the path from the reference element of the radar transmitter to the reference element of the forward \gls{ris};
	
	\item $\bar{\delta}_{jn}$, $\bar{\rho}_{jn}^{\rm az}$, $\bar{\rho}_{jn}^{\rm el}$, $\bar{\omega}_{jn}^{\rm az}$, and $\bar{\omega}_{jn}^{\rm el}$ are the length, the azimuth and elevation angles of departure, and the azimuth and elevation angles of arrival, respectively, of the path  from the $j$-th element of the radar transmitter to the $n$-th element of the forward \gls{ris}, respectively; clearly, $\bar{\delta}_{jn}=\bar{\delta}$, $\bar{\rho}_{jn}^{\rm az}=\bar{\rho}^{\rm az}$,  $\bar{\rho}_{jn}^{\rm el}=\bar{\rho}^{\rm el}$, $\bar{\omega}_{jn}^{\rm az}=\bar{\omega}^{\rm az}$, and $\bar{\omega}_{jn}^{\rm el}=\bar{\omega}^{\rm el}$ if the $j$-th transmit and the $n$-th forward reflecting element are the reference one of the corresponding array.
\end{itemize}
Similar parameters are defined at the receive side, with an obvious modification of the notation, i.e., the diacritical mark $\bar{(\cdot)}$ is replaced by $\ddot{(\cdot)}$.  For brevity, we denote by $\phi=\{\phi^{\rm az},\phi^{\rm el}\}$ the azimuth and elevation angles $\phi^{\rm az}$ and $\phi^{\rm el}$. These parameters are summarized in Fig.~\ref{fig:system}.

\subsection{Design assumptions}
At the design stage, we make the following assumptions.
\begin{itemize}
	\item There is no coupling among the elements of the considered transmit/receive/reflecting arrays.	
	
	\item There is only \gls{los} propagation in the radar-\gls{ris}, radar-target, and \gls{ris}-target hops (whenever present).
	
	\item Any pair of transmit and forward reflecting elements are in each other's far-field; similarly, any pair of receive and backward reflecting elements are in each other's far-field.  This requires that~\cite{book-Stutzman03}
	\begin{subequations}\label{delta_min}
		\begin{align}
			\min_{j,n}\bar{\delta}_{jn}& \geq \max\{2\bar{\Delta}_{r}^{2}/\lambda,2\bar{\Delta}_{s}^{2}/\lambda,5\bar{\Delta}_{r},5\bar{\Delta}_{s},1.6\lambda\}\\
			\min_{j,n}\ddot{\delta}_{jn}& \geq \max\{2\ddot{\Delta}_{r}^{2}/\lambda,2\ddot{\Delta}_{s}^{2}/\lambda,5\ddot{\Delta}_{r},5\ddot{\Delta}_{s},1.6\lambda\}
		\end{align}
	\end{subequations}
	where $\lambda$ is the carrier wavelength and $\bar{\Delta}_{r}$,  $\bar{\Delta}_{s}$, $\ddot{\Delta}_{r}$, and $\ddot{\Delta}_{s}$ are the maximum size of each element of the transmit, forward reflecting, receive,  and backward reflecting arrays, respectively. We underline that the radar transmitter and the forward \gls{ris} and, similarly,  the radar receiver and the backward \gls{ris} are not required to be in each other's far-field. Accordingly, the phase curvature of the wavefront across the array elements is not neglected in the radar-\gls{ris} hops~\cite{Friedlander-2019,2021-Tulino}.
	
	\item The target  and  any radar array are in each other's far-field, i.e., we have~\cite{book-Stutzman03}
	\begin{subequations}
		\begin{align}
			\bar{d}_{r}&\geq \max\{2\bar{D}_{t}^{2}/\lambda,2\bar{D}_{r}^{2}/\lambda,5\bar{D}_{t},5\bar{D}_{r},1.6\lambda\}\\
			\ddot{d}_{r}&\geq \max\{2\ddot{D}_{t}^{2}/\lambda,2\ddot{D}_{r}^{2}/\lambda,5\ddot{D}_{t},5\ddot{D}_{r},1.6\lambda\}
		\end{align}
	\end{subequations}
	where $\bar{D}_{r}$ and $\ddot{D}_{r}$ are the maximum size of the transmit and the receive array, respectively, and $\bar{D}_{t}$  and $\ddot{D}_{t}$ are the effective size of the target as seen from the radar transmitter and receiver, respectively; also, the target and any \gls{ris} are in each other's far-field, i.e., we have~\cite{book-Stutzman03}
	\begin{subequations}
		\begin{align}
			\bar{d}_{s}&\geq \max\{2\bar{D}_{t}^{2}/\lambda,2\bar{D}_{s}^{2}/\lambda,5 \bar{D}_{t}, 5\bar{D}_{s},1.6\lambda\}\\
			\ddot{d}_{s}&\geq \max\{2\ddot{D}_{t}^{2}/\lambda,2\ddot{D}_{s}^{2}/\lambda, 5\ddot{D}_{t},5\ddot{D}_{s},1.6\lambda\}
		\end{align}
	\end{subequations}
	where $\bar{D}_{s}$ and $\ddot{D}_{s}$ are the maximum size of the forward and backward \glspl{ris}, respectively. Accordingly,  the phase curvature of the wavefront is neglected along the radar-target and  \gls{ris}-target hops~\cite{book-Stutzman03,book-Balanis}.		
	
	\item When both the radar transmitter and the forward \gls{ris} illuminate the prospective target, they see the same aspect angle~\cite{2006-Fishler-MIMOradar}, i.e.,  $\lambda/\bar{D}_{t}\gg\max_{j,n}\bar{\xi}_{jn}$, where  $\bar{\xi}_{jn}$ is the angle formed by the line segment linking the target with the $j$-th transmit element and the line segment linking the target with the $n$-th forward reflecting element; similarly, when both the radar receiver and the backward \gls{ris} observe the prospective target, they see the same aspect angle, i.e., $\lambda/\ddot{D}_{t}\gg \max_{j,n}\ddot{\xi}_{jn}$, where $\ddot{\xi}_{jn}$ is the angle formed by the line segment linking the target with the $j$-th receive element and the line segment linking the target with the $n$-th backward reflecting element.	
    \item The waveforms are narrowband, so that the delays of the target echoes reaching the  receiver are not resolvable.     
\end{itemize}

\subsection{Received signal}\label{Sec:rx-signal}
A range gating operation is at first performed by projecting the signal impinging on each receive element along a delayed version of each transmit waveform (this is tantamount to sampling the output of a filter matched to the waveform), with the delay tied to the distance of the inspected resolution cell from the transmitter and the receiver~\cite{2006-Fishler-MIMOradar,book-Li-Stoica}. The available data samples are organized into a vector $\bm{r}\in\mathbb{C}^{\ddot{N}_{r}\bar{N}_{r}}$ that, in the presence of a steady target, can be modeled as 
\begin{equation}\label{vector-model}
	\bm{r}=\bm{e}\alpha+\bm{w}
\end{equation}
where  $\bm{e}\in\mathbb{C}^{\ddot{N}_{r}\bar{N}_{r}}$ is the known  \emph{target signature}, tied to its location, the system geometry, and the  phase shifts of the \glspl{ris} (more on this infra), $\alpha\in\mathbb{C}$ accounts for the unknown target response and any other scaling factor not included in the target signature (such as, for example, the transmit power); $\bm{w}\in\mathbb{C}^{\ddot{N}_{r}\bar{N}_{r}}$ is a circularity-symmetric Gaussian vector with covariance matrix $\sigma_{w}^{2}\bm{I}_{N_{r}}$, accounting for the additive noise.\footnote{The following developments can also  be extended to the case where the covariance matrix of $\bm{w}$ is full-rank and has the separable structure $\text{E}[\bm{w}\bm{w}^{H}]=\bar{\bm{C}}_{w}\otimes  \ddot{\bm{C}}_{w}$, with  $\bar{\bm{C}}_{w}\in \mathbb{C}^{\bar{N}_{r}\times \bar{N}_{r}}$ and $\ddot{\bm{C}}_{w}\in \mathbb{C}^{\ddot{N}_{r}\times \ddot{N}_{r}}$.}

The target signature contains the superposition of up to four paths from the radar transmitter to the target to the radar receiver (see Fig.~\ref{fig:system}), which may involve up to two bounces off the \glspl{ris}. Let $\bar{\bm{x}}\in\mathbb{C}^{\bar{N}_{s}}$ and  $\ddot{\bm{x}}\in\mathbb{C}^{\ddot{N}_{s}}$ be the vectors with unit modulus entries specifying the phase shifts introduced by the elements of the forward and backward \glspl{ris}, respectively. Then, we have
\begin{align}\label{Kronecker-structure}
	&\bm{e}(\bar{\bm{x}},\ddot{\bm{x}})= \underbrace{ \big( \bar{\bm{v}}_{r}\bar{\gamma}_{r}\big)\otimes \big( \ddot{\bm{v}}_{r}\ddot{\gamma}_{r}\big)}_{\text{Radar$\rightarrow$Target$\rightarrow$Radar}}\notag \\	
	&\quad +  \underbrace{\big( \bar{\bm{G}}\diag\bigl\{\bar{\bm{x}}\bigr\}\bar{\bm{v}}_{s} \bar{\gamma}_{s}\big)\otimes \big(  \ddot{\bm{v}}_{r}\ddot{\gamma}_{r}\big) }_{\text{Radar$\rightarrow$RIS$\rightarrow$Target$\rightarrow$Radar}} \notag \\	
	 &\quad + \underbrace{\big( \bar{\bm{v}}_{r}\bar{\gamma}_{r}\big)\otimes \big( \ddot{\bm{G}} \diag\bigl\{\ddot{\bm{x}}\bigr\} \ddot{\bm{v}}_{s}\ddot{\gamma}_{s}\big) }_{\text{Radar$\rightarrow$Target$\rightarrow$RIS$\rightarrow$Radar}} \notag\\
	&\quad +\underbrace{\big( \bar{\bm{G}}\diag\bigl\{\bar{\bm{x}}\bigr\}\bar{\bm{v}}_{s} \bar{\gamma}_{s}\big)\otimes \big( \ddot{\bm{G}} \diag\bigl\{\ddot{\bm{x}}\bigr\} \ddot{\bm{v}}_{s}\ddot{\gamma}_{s}\big) }_{\text{Radar$\rightarrow$RIS$\rightarrow$Target$\rightarrow$RIS$\rightarrow$Radar}}\notag\\
	&=\underbrace{\bigl(\bar{\bm{v}}_{r}\bar{\gamma}_{r}+\bar{\bm{G}}\diag\bigl\{\bar{\bm{x}}\bigr\}\bar{\bm{v}}_{s}\bar{\gamma}_{s} \bigr)}_{\substack{\bar{\bm{e}}(\bar{\bm{x}})\\ \text{transmit target signature}}} \otimes \underbrace{\bigl(\ddot{\bm{v}}_{r}\ddot{\gamma}_{r}+\ddot{\bm{G}} \diag\bigl\{\ddot{\bm{x}}\bigr\} \ddot{\bm{v}}_{s} \ddot{\gamma}_{s}\bigr)}_{\substack{\ddot{\bm{e}}(\ddot{\bm{x}}) \\ \text{receive target signature}} }
\end{align}
where we have made explicit the dependence of the vector $\bm{e}$ upon $\bar{\bm{x}}$ and $\ddot{\bm{x}}$ for future developments and we have defined the symbols listed below.
\begin{itemize}		
	\item $\bar{\bm{v}}_{r}\in\mathbb{C}^{\bar{N}_r}$ and $\ddot{\bm{v}}_{r}\in\mathbb{C}^{\ddot{N}_r}$ are the \emph{direct} steering vectors of the radar transmitter and receiver towards the directions $\bar{\theta}_r$ and $\ddot{\theta}_r$ of the target, respectively, which are tied to the array geometry and have unit modulus entries.
	
	\item  $\bar{\bm{v}}_{s}\in\mathbb{C}^{\bar{N}_s}$ and $\ddot{\bm{v}}_{s}\in\mathbb{C}^{\ddot{N}_s}$ are the steering vectors of the forward and backward \glspl{ris} towards the directions $\bar{\theta}_s$ and $\ddot{\theta}_s$ of the target, respectively, which are tied to the array geometry and have unit modulus entries.
	
	\item 	$\bar{\gamma}_{r}\in\mathbb{C}$ and $\ddot{\gamma}_{r}\in\mathbb{C}$ are the direct channels between the reference transmit element and the target and between the target and the reference receive element, respectively, which account for the radar element gain, the path-loss, and the phase delay.

	\item $\bar{\gamma}_{s}\in\mathbb{C}$ and $\ddot{\gamma}_{s}\in\mathbb{C}$ are the two-hop indirect channels from the reference transmit element to the reference forward reflecting element to the target and from the target to the reference backward reflecting element to the reference receive element, respectively, which account for the radar element gain, the bistatic \gls{rcs} of the reflecting element, and the two-hop path-loss and phase delay. Hereafter, we assume that $\bar{\gamma}_{s}=0$ if the forward \gls{ris} is not present; similarly, $\ddot{\gamma}_{s}=0$ if the backward \gls{ris} is not present.  
	
	\item  $\bar{\bm{G}}\in \mathbb{C}^{ \bar{N}_{r} \times \bar{N}_{s}}$ and $\ddot{\bm{G}}\in \mathbb{C}^{\ddot{N}_{r}\times \ddot{N}_{s}}$ are the normalized channel matrices between the radar transmitter and the forward \gls{ris} and between the backward \gls{ris} and the radar receiver, respectively, whose entries account for the radar element gain, the bistatic \gls{rcs} of the reflecting element, the path-loss, and the phase delay; the normalization is with respect to the scalar channel between the reference elements of the arrays that is included in $\bar{\gamma}_{s}$ and $\ddot{\gamma}_{s}$. 	
\end{itemize}
Depending on the values of $\bar{\gamma}_{r}$,  $\ddot{\gamma}_{r}$, $\bar{\gamma}_{s}$,  and  $\ddot{\gamma}_{s}$, only some of the echoes in~\eqref{Kronecker-structure} are actually present: the possible system configurations are outlined in Table~\ref{table:configurations}. Our model subsumes the one considered in~\cite{MIMO_sensors}, wherein the radar has a \gls{los}-view of the prospective target (i.e., $\bar{\gamma}_{r}\ddot{\gamma}_{r}\neq0$) and only uses a backward \gls{ris} (i.e., $\bar{\gamma}_{s}=0$ and $\ddot{\gamma}_{s}\neq0$).  Also, notice that  $\bm{e}(\bar{\bm{x}},\ddot{\bm{x}})$  possesses a Kronecker structure, wherein  $\bar{\bm{G}}\diag\bigl\{\bar{\bm{x}}\bigr\} \bar{\bm{v}}_{s}$ and $\ddot{\bm{G}} \diag\bigl\{\ddot{\bm{x}}\bigr\} \ddot{\bm{v}}_{s}$ are the \emph{indirect} steering vectors of the radar  transmitter  and receiver towards the target via the \gls{ris}-assisted path.\footnote{We use here the term ``steering vector'' with some abuse of notation as the entries of  $\bar{\bm{G}}\diag\bigl\{\bar{\bm{x}}\bigr\} \bar{\bm{v}}_{s}$ and $\ddot{\bm{G}} \diag\bigl\{\ddot{\bm{x}}\bigr\} \ddot{\bm{v}}_{s}$ may not have a unit modulus.} Accordingly, $\bar{\bm{e}}(\bar{\bm{x}})$ is the \emph{transmit target signature}, resulting from the superposition of the direct and indirect steering vectors of the radar transmitter towards the target with weights equal to the corresponding channels $\bar{\gamma}_{r}$ and $\bar{\gamma}_{s}$, while $\ddot{\bm{e}}(\ddot{\bm{x}})$ is the \emph{receive target signature},  resulting  from the superposition of the direct and indirect steering vectors  of the radar receiver towards the target with weights equal to the corresponding channels $\ddot{\gamma}_{r}$ and $\ddot{\gamma}_{s}$.

\begin{table}[!t]
	\centering
	\caption{Possible system configurations\label{table:configurations}}	
	\begin{tabular}{lllllll}
		\toprule 
		\multirow{2}[2]{*}{Radar} & \multirow{2}[2]{*}{$\bar{\gamma}_{r}$} & \multirow{2}[2]{*}{$\ddot{\gamma}_{r}$} & \multirow{2}[2]{*}{$\bar{\gamma}_{s}$} & \multirow{2}[2]{*}{$\ddot{\gamma}_{s}$} & \multicolumn{2}{c}{Target} \\
		
		\cmidrule(rl){6-7}  & & & & &  illuminated by & observed by    \\
		
		\midrule
		\multirow{4}{*}{LOS} & \multirow{4}{*}{$\neq0$} & \multirow{4}{*}{$\neq0$} & \cc $\neq0$ & \cc $\neq0$ & \cc Radar \& RIS & \cc Radar \& RIS   \\
		
		& &  & $=0$ & $\neq0$ &  Radar & Radar \& RIS    \\
		
		& &  & \cc $\neq0$ & \cc $=0$ &  \cc Radar \& RIS & \cc Radar    \\
		
		& &  & $=0$ & $=0$ &  Radar  & Radar   \\
		
		\cmidrule(rl){1-7}		 
		\multirow{5}{*}[-5pt]{NLOS} & \multirow{2}{*}{$\neq0$} & \multirow{2}{*}{$=0$}& \cc $\neq0$ & \cc $\neq0$ & \cc Radar \& RIS & \cc \hspace{0.99cm} RIS \\
		
		& & & $=0$ & $\neq0$ & Radar & \hspace{0.99cm} RIS    \\
		
		\cmidrule(rl){2-7}

& $=0$ & $=0$ & \cc $\neq0$ & \cc  $\neq0$ & \cc \hspace{0.99cm} RIS &\cc  \hspace{0.99cm} RIS \\
		
		\cmidrule(rl){2-7}
		
		& \multirow{2}{*}{$=0$} & \multirow{2}{*}{$\neq0$} &  $\neq0$ &  $\neq0$ &  \hspace{0.99cm} RIS &  Radar \& RIS    \\
		
		& &  & \cc $\neq0$ & \cc $=0$ & \cc \hspace{0.99cm} RIS & \cc Radar   \\

		\bottomrule			
	\end{tabular}
\end{table}

\subsection{Channel model}\label{Sec:channel}
The design methodology proposed in the following section is independent of the model  adopted for the radar-target and the radar-RIS-target  channels (both ways), as long  as  the received signal can be written as in~\eqref{vector-model} and \eqref{Kronecker-structure}. Exact modeling of an    \gls{ris}-assisted channel is a non-trivial and still debated problem. In order to shed some light on the potential advantages granted by the \glspl{ris} in the \gls{mimo} radar target detection, at the analysis stage we just leverage and adapt basic models so far elaborated for RIS-aided communications.

According to the standard radar equation~\cite{book-Meyer,book-Richards},  the scalar channels $\bar{\gamma}_{r}$, $\bar{\gamma}_{s}$, $\ddot{\gamma}_{r}$, and $\ddot{\gamma}_{s}$ are modeled as
\begin{subequations}\label{gamma-def}
	\begin{align}
		\bar{\gamma}_{r}&=\left(\frac{\bar{\mathcal{G}}(\bar{\theta}_{r})}{4\pi\bar{d}_{r}^2\bar{L}_{r}}\right)^{1/2} e^{-\i 2 \pi \bar{d}_{r}/\lambda} \\	
		\ddot{\gamma}_{r}&=\left(
		\frac{ \ddot{\mathcal{G}}(\ddot{\theta}_{r})\lambda^2}{(4\pi)^{2} \ddot{d}_{r}^2\ddot{L}_{r}}\right)^{1/2} e^{-\i 2 \pi\ddot{d}_{r}/\lambda} \\	\bar{\gamma}_{s}&= \left(\frac{\bar{\mathcal{G}}(\bar{\rho}) \bar{\zeta}(\bar{\omega},\bar{\theta}_{s})}{(4\pi)^{2}  \bar{\delta}^2\bar{d}_{s}^2 \bar{L}_{s} }\right)^{1/2}  e^{-\i 2 \pi (\bar{\delta}+\bar{d}_{s})/\lambda}  \\	
		\ddot{\gamma}_{s}&= \left(\frac{  \ddot{\zeta}(\ddot{\theta}_{s},\ddot{\omega}) \ddot{\mathcal{G}}(\ddot{\rho})\lambda^2}{(4\pi)^{3}    \ddot{d}_{s}^2 \ddot{\delta}^2 \ddot{L}_{s}}\right)^{1/2} e^{-\i 2 \pi (\ddot{d}_{s}+\ddot{\delta})/\lambda}
	\end{align}
\end{subequations}
where $\bar{\mathcal{G}}(\varphi)$ and $\ddot{\mathcal{G}}(\varphi)$ are the gain of the transmit and receive elements in the direction $\varphi=\{\varphi^{\rm az}, \varphi^{\rm el}\}$, respectively, $\bar{\zeta}(\varphi_{\rm in}, \varphi_{\rm out})$  and $\ddot{\zeta}(\varphi_{\rm in}, \varphi_{\rm out})$ are the bistatic \gls{rcs} of the forward and backward reflecting elements towards the direction $\varphi_{\rm out}=\{\varphi_{\rm out}^{\rm az},\varphi_{\rm out}^{\rm el}\}$ when illuminated from the direction $\varphi_{\rm in}=\{\varphi_{\rm in}^{\rm az},\varphi_{\rm in}^{\rm el}\}$, respectively, and  $\bar{L}_{r}$, $\ddot{L}_{r}$, $\bar{L}_{s}$,  and $\ddot{L}_{s}$ are loss factors accounting for any additional attenuation along the corresponding paths, respectively.

Following~\cite{Friedlander-2019,2021-Tulino}, the entries of channel matrix $\bar{\bm{G}}$ 
are modeled as  $\bar{G}_{jn}=0$, if $\bar{\gamma}_{s}=0$, and	 
\begin{equation}\label{Gbar-def}
	\bar{G}_{jn}=\left(\frac{\bar{\mathcal{G}}(\bar{\rho}_{jn}) \bar{\zeta}(\bar{\omega}_{jn},\bar{\theta}_{s}) \bar{\delta}^2}{\bar{\mathcal{G}}(\bar{\rho}) \bar{\zeta}(\bar{\omega},\bar{\theta}_{s})\bar{\delta}_{jn}^2}\right)^{1/2}
	e^{-\i 2\pi (\bar{\delta}_{jn}-\bar{\delta})/ \lambda} 
\end{equation}
otherwise. In this latter case, $\bar{G}_{jn}=1$, if the $j$-th transmit and the $n$-th forward reflecting element are the reference ones of the corresponding arrays, in keeping with~\eqref{gamma-def}. Similarly, the entries of $\ddot{\bm{G}}$ are modeled as $\ddot{G}_{jn}=0$,  if $\ddot{\gamma}_{s}=0$, and
\begin{equation}\label{Gddot-def}
	\ddot{G}_{jn}=\left(\frac{\ddot{\zeta}(\ddot{\theta}_{s},\ddot{\omega}_{jn}) \ddot{\mathcal{G}}(\ddot{\rho}_{jn})  \ddot{\delta}^2}{ \ddot{\zeta}(\ddot{\theta}_{s},\ddot{\omega})\ddot{\mathcal{G}}(\ddot{\rho})\ddot{\delta}_{jn}^2}\right)^{1/2}
e^{-\i 2\pi (\ddot{\delta}_{jn}-\ddot{\delta})/ \lambda} 
\end{equation}
otherwise.

Finally, leveraging~\cite{Ellingson2019ris,DiRenzo2021-modeling-pathloss,Najafi-2021,2021-Tulino},  the bistatic \gls{rcs} of a forward reflecting element is modeled as
\begin{multline}
	\bar{\zeta}(\varphi_{\rm in}, \varphi_{\rm out})= \underbrace{\bar{A}[\cos(\varphi_{\rm in}^{\rm az})\cos(\varphi_{\rm in}^{\rm el})]^{q}}_{\bar{\zeta}_{a}(\varphi_{\rm in})}\\\times \underbrace{(4 \pi \bar{A}/\lambda^2)[\cos(\varphi_{\rm out}^{\rm az})\cos(\varphi_{\rm out}^{\rm el})]^{q}}_{\bar{\zeta}_{g}(\varphi_{\rm out})}
	\label{bistatic_rcs}
\end{multline}
if $\varphi_{\rm in}^{\rm az},\varphi_{\rm in}^{\rm el},\varphi_{\rm out}^{\rm az},\varphi_{\rm out}^{\rm el}\in(-\pi/2,\pi/2)$ and $\bar{\zeta}(\varphi_{\rm in}, \varphi_{\rm out})=0$ otherwise, with $\bar{A}$ being the area of the element. This is a simple yet realistic model, wherein the \gls{rcs} is regarded as the product of an effective receive aperture $\bar{\zeta}_{a}(\varphi_{\rm in})$  and a transmit gain $\bar{\zeta}_{g}(\varphi_{\rm out})$, with a cosine-shaped scan loss in both azimuth and elevation. In order to conserve the power, the cosine exponent $q\geq0$ must be such that the integral of $\bar{\zeta}_{g}(\varphi_{\rm out})$ over a sphere does not exceed $4\pi$ sr. This model treats the \gls{ris} as a reciprocal surface, i.e., $\bar{\zeta}(\varphi_{\rm in}, \varphi_{\rm out})=\bar{\zeta}(\varphi_{\rm out}, \varphi_{\rm in})$, and, at broadside, the entire reflecting area is equal to the sum of the effective apertures of the building elements, so that the overall \gls{ris} behaves as a flat plate of area $\bar{N}_{s}\bar{A}$~\cite{book-Balanis}. The model in~\eqref{bistatic_rcs}  is also used  for $\ddot{\zeta}(\varphi_{\rm in}, \varphi_{\rm out})$ upon replacing $\bar{A}$ with the area  $\ddot{A}$ of a backward reflecting element.

 \section{System design}\label{SEC:System-design}
Let $\bm{f}\in\mathbb{C}^{\ddot{N}_{r}\bar{N}_{r}}$ be the unit-norm filter employed by the receiver to focus the radar towards the location under inspection, which is under the designer's control; then, its output is 
\begin{equation}\label{output-filter-u}
	r= \bm{f}^{\herm} \bm{e}(\bar{\bm{x}},\ddot{\bm{x}})\alpha+\bm{f}^{\herm}\bm{w}
\end{equation} 
resulting in the following \gls{snr} 
\begin{equation}\label{output-snr}
	\text{SNR}= \left|\bm{f}^{\herm}  \bm{e}(\bar{\bm{x}},\ddot{\bm{x}}) \right|^2\frac{\sigma_{\alpha}^2}{\sigma^2_{w}}
\end{equation}
where $\sigma_{\alpha}^2=\text{E}[|\alpha|^2]$. Also, the \gls{glrt} discriminating between the target presence (hypothesis $H_{1}$) and absence  (hypothesis $H_{0}$)  is~\cite{KayBook_vol2}
\begin{equation}
	\frac{\left|r\right|^2}{\sigma^2_{w}}	\test \eta
\end{equation}
where $\eta>0$ is the detection threshold, to be set according to the desired probability of false alarm  $\text{P}_{\text{fa}}=e^{-\eta}$; assuming that $|\alpha|^2$ is non fluctuating, exponentially distributed, or gamma distributed with variance $\sigma^{2}_{\alpha}/2$,  the corresponding detection probabilities are~\cite{book-Meyer,book-Richards}
\begin{equation}\label{Pd_coherent}
	\text{P}_{\text{d}}= \begin{cases}
		Q_{1}\left(\sqrt{2 \text{SNR}},\sqrt{ 2 \eta}\right)&  \text{(non fluctuating)}\\
		e^{-\gamma/(1+\text{SNR})} & \text{(exponential)}\\
		\left( 1+ \frac{\gamma \text{SNR}/2}{\left(1+\text{SNR}/2\right)^2}\right)
		e^{-\gamma/(1+\text{SNR}/2)} & \text{(gamma)}.\\
	\end{cases}
\end{equation}
where $Q_1(\cdot, \cdot)$ is the Marcum's $Q-$function. 

The receive filter and the phase shifts of the \gls{ris} can be jointly  chosen to maximize the \gls{snr} at the location under inspection and, as a by product, the detection probability in~\eqref{Pd_coherent}. For any $\bar{\bm{x}}$ and $\ddot{\bm{x}}$, the optimal filter is the one matched to the signal to be detected, i.e., 
\begin{equation}\label{opt-filter}
	\bm{f}=\frac{\bm{e}(\bar{\bm{x}},\ddot{\bm{x}})}{\bigl\|\bm{e}(\bar{\bm{x}},\ddot{\bm{x}})\bigr\|}.
\end{equation}
Hence, upon plugging~\eqref{opt-filter} into~\eqref{output-snr}, the optimal phase shifts are obtained as the solution to 
\begin{equation}\label{problem-coherent-2}
  \begin{aligned}
\max_{\bar{\bm{x}},\,\ddot{\bm{x}}}&\quad\|\bm{e}(\bar{\bm{x}},\ddot{\bm{x}}) \|^2\\
  \text{s.t.}& \quad|\bar{{x}}_{n}|=1,\;n=1,\ldots,\bar{N}_{s}\\
             & \quad |\ddot{{x}}_{n}|=1,\;n=1,\ldots,\ddot{N}_{s}
  \end{aligned}
\end{equation}
which is independent of the target and noise strength (namely, of $\sigma_{\alpha}^{2}$ and $\sigma_{w}^{2}$). Notice that the design in~\eqref{problem-coherent-2} can be regarded as a form of passive beamforming. Indeed, by acting on the response of the forward and backward reflecting elements, the indirect steering vectors of the radar transmitter and receiver towards the target can be modified to provide an SNR gain, respectively; such gain results from the constructive alignment of the indirect signals reflected by each element of the RIS and of the direct signal during both the illumination and the observation of the target. Since the objective function $\|\bm{e}(\bar{\bm{x}},\ddot{\bm{x}})\|^2=\|\bar{\bm{e}}(\bar{\bm{x}})\|^2\|\ddot{\bm{e}}(\ddot{\bm{x}})\|^2$ and the constraint set in~\eqref{problem-coherent-2} are separable with respect to the variables $\bar{\bm{x}}$ and $\ddot{\bm{x}}$, the forward and backward phase shifts can be optimized independently; specifically, the problems to be solved are 
\begin{equation}\label{problem-PHI-design}
	\max_{\bar{\bm{x}}}\|\bar{\bm{e}}(\bar{\bm{x}})\|^2,\quad
	  \text{s.t. } |\bar{{x}}_{n}|=1,\;n=1,\ldots,\bar{N}_{s}
\end{equation}  
when the forward \gls{ris} is present, and
\begin{equation}\label{problem-barPHI-design}
\max_{\ddot{\bm{x}}}\|\ddot{\bm{e}}(\ddot{\bm{x}})\|^2,\quad
  \text{s.t. } |\ddot{{x}}_{n}|=1,\;n=1,\ldots,\ddot{N}_{s}
\end{equation}  
when the backward \gls{ris} is present. 

\subsection{Optimization of the phase shifts of the \gls{ris}}\label{Sec:disjoint-phase-design}
We discuss here the solution to~\eqref{problem-PHI-design}, while similar arguments can be used to tackle~\eqref{problem-barPHI-design}. Let  $\bar{\bm{y}}=\bar{\bm{v}}_{r}\bar{\gamma}_{r}$ and $\bar{\bm{Q}}= \bar{\bm{G}}\diag\left(\bar{\bm{v}}_{s}\right)\bar{\gamma}_{s}$, which are computed based only on the knowledge of the system geometry (i.e., the position of the radar transmit elements and the forward-reflecting elements, which are available), the location to be inspected (which is decided by the radar engineer), the radar-\gls{ris} channel (which can be estimated in advance), and the far-field path-loss model (which can be obtained from experimental data and/or theoretical models). Problem~\eqref{problem-PHI-design} is now rewritten as
\begin{equation}\label{problem-PHI-design-2}
	\max_{\bar{\bm{x}}}\; \bigl\| \bar{\bm{y}}+ \bar{\bm{Q}} \bar{\bm{x}} \bigr\|^2,\quad
	\text{s.t. } |\bar{{x}}_{n}|=1,\;n=1,\ldots,\bar{N}_{s}.
\end{equation}

When $\rank(\bar{\bm{Q}})=1$, which occurs if the radar transmitter and the forward \gls{ris} are in each other's far-field (more on this in Sec.~\ref{SEC:Far-field}) or $\bar{N}_{r}=1$ or $\bar{N}_{s}=1$, then the solution to~\eqref{problem-PHI-design-2} is derived in closed form. Indeed, upon factorizing $\bar{\bm{Q}}$ as  $\bar{\bm{a}}\bar{\bm{b}}^{T}$, with $\bar{\bm{a}}\in\mathbb{C}^{\bar{N}_{r}}$ and $\bar{\bm{b}}\in\mathbb{C}^{\bar{N}_{s}}$, we have that 
\begin{align}
\bigl\|\bar{\bm{y}}+ \bar{\bm{a}}\bar{\bm{b}}^{\transp} \bar{\bm{x}}\bigr\|^{2} 
	&= \|\bar{\bm{y}}\|^2+\|\bar{\bm{a}}\|^2 \Biggl|\sum_{n=1}^{\bar{N}_{s}}\bar{b}_{n}\bar{x}_{n} \Biggr|^2\notag \\
	&\quad +2\Re\Biggl\{\bar{\bm{y}}^{\herm}\bar{\bm{a}}\sum_{n=1}^{\bar{N}_{s}}\bar{b}_{n}\bar{x}_{n}\Biggr\}\notag\\
	&\leq \|\bar{\bm{y}}\|^2+\|\bar{\bm{a}}\|^2 \Biggl(\sum_{n=1}^{\bar{N}_{s}}|\bar{b}_{n}| \Biggr)^2\!\!\! +2|\bar{\bm{y}}^{\herm}\bar{\bm{a}}|\sum_{n=1}^{\bar{N}_{s}}|\bar{b}_{n}|\label{phase-design-rank-1}
\end{align} 
where the last upper bound is achieved when 
\begin{equation}\label{phase-design-rank-1-sol}
	\angle \bar{x}_{n}=-\angle \bar{b}_{n} - \angle (\bar{\bm{y}}^{\herm}\bar{\bm{a}}), \quad n=1,\ldots, \bar{N}_{s}.
	\end{equation}

When $\rank(\bar{\bm{Q}})>1$, upon introducing the following positive semi-definite matrix\footnote{The positive-semidefiniteness is verified by exploiting the Schur complement for block matrices.}
\begin{align}\label{B-matrix}
 \bar{\bm{B}} =\begin{pmatrix} \bar{\bm{Q}}^{\herm}\bar{\bm{Q}} &&& \bar{\bm{Q}}^{\herm}\bar{\bm{y}}\\
 \bar{\bm{y}}^{\herm} \bar{\bm{Q}} &&& \Vert \bar{\bm{y}} \Vert^2\end{pmatrix} \in \mathbb{C}^{(\bar{N}_{s}+1) \times (\bar{N}_{s}+1)}
\end{align}
Problem~\eqref{problem-PHI-design-2} is recast as
\begin{equation}
\begin{aligned}
	\max_{\bar{\bm{x}}, \phi} & \quad \bigl(\e^{-\i \phi} \bar{\bm{x}}^{\herm} \;\; \e^{-\i \phi} \bigr) \bar{\bm{B}} \bigl(\e^{\i \phi}\bar{\bm{x}}^{\transp} \;\; \e^{\i \phi}\bigr)^{\transp}\\
	\text{s.t. }& \quad |\bar{{x}}_{n}|=1,\;n=1,\ldots,\bar{N}_{s}
\end{aligned}
\end{equation}
which in turn is equivalent to
\begin{equation}\label{problem-x-design_2}
		\max_{\bar{\bm{z}}} \; \bar{\bm{z}}^{\herm} \bar{\bm{B}} \bar{\bm{z}},\quad	\text{s.t. }  |\bar{z}_n|=1,\; n=1,\ldots, \bar{N}_{s}+1.
\end{equation}
If $\bar{\bm{z}}^\star$ solves~\eqref{problem-x-design_2}, the solution to~\eqref{problem-PHI-design-2} can be recovered as $\angle \bar{x}_n= \angle \bar{z}_n^\star- \angle \bar{z}_{\bar{N}_{s}+1}^\star$, $n=1,\ldots,\bar{N}_{s}$. Problem~\eqref{problem-x-design_2} is a complex quadratic program, that is strongly NP-hard in general~\cite{Zhang_Huang_2006, So_Zhang_Ye_2007}. A sub-optimal solution can be obtained via an alternate maximization, as reported in Algorithm~\ref{alg_x_design}~\cite{Waldspurger_2015}. In this case, one entry of $\bar{\bm{z}}$ at a time is iteratively optimized; since $|\bar z_n|=1$ for any $n$, the objective function of~\eqref{problem-x-design_2} can be written as $2\Re \bigl\{ \bar{z}_n^* \sum_{j\neq n} \bar{B}_{nj} \bar{z}_j\bigr\} + \bar{B}_{nn}+ \sum_{k\neq n} \sum_{j\neq n} \bar{z}_k^* \bar{B}_{kj} \bar{z}_j$, and the constrained maximization over $\bar{z}_{n}$ gives
\begin{equation}
 \bar{z}_n=\e^{\i\angle \left(\sum_{j\neq n} \bar{B}_{nj} \bar{z}_j\right)}
\end{equation}
if $\sum_{j\neq n} \bar{B}_{nj} \bar{z}_j\neq0$, and any unit modulus complex number, otherwise. Since the objective function is bounded above and monotonically increased at each iteration, convergence is ensured. The complexity per iteration is $\mathcal O(\bar{N}_{s}^2)$, with an initial cost of $\mathcal O(\bar{N}_{s}^3)$ to form the matrix $\bar{\bm{B}}$ in~\eqref{B-matrix}.

\begin{algorithm}[t]
	\caption{Alternate maximization for Problem~\eqref{problem-x-design_2}}
	\begin{algorithmic}[1]\label{alg_x_design}
		\STATE Choose $\epsilon>0$, $K _\text{max}>0$, and $\bar{\bm{z}}\in \mathbb{C}^{\bar{N}_{s}+1}$: $|\bar{z}_{n}|=1$, $\forall n$ 
		\STATE $k=0$ and $\bar{f}_{0}= \bar{\bm{z}}^{\herm} \bar{\bm{B}} \bar{\bm{z}}$
		\REPEAT
		\FOR{$n=1,\ldots,N_{s}+1$}
		\STATE $\bar{z}_{n}= \begin{cases} \e^{\i\angle \left(\sum_{j\neq n} \bar{B}_{nj} \bar{z}_j\right)}, & \text{if } \sum_{j\neq n} \bar{B}_{nj} \bar{z}_j\neq 0\\
		1, & \text{otherwise}
		\end{cases}$
		\ENDFOR
		\STATE $k = k+1$
		\STATE $\bar{f}_{k}=\bar{\bm{z}}^{\herm} \bar{\bm{B}} \bar{\bm{z}}$
		\UNTIL{$\bar{f}_{k}-\bar{f}_{k-1}<\epsilon \bar{f}_{k}$ or $k=K_\text{max}$}
	\end{algorithmic}
\end{algorithm}

Alternatively (see, e.g.,~\cite{Goemans_1995, Goemans_2004, Zhang_Huang_2006, So_Zhang_Ye_2007, Luo_2010}), Problem~\eqref{problem-x-design_2} can be reformulated as
\begin{equation}
	\begin{aligned}\label{problem-x-design_3}
		\max_{\bar{\bm{Z}}, \bar{\bm{z}}} &\quad \trace( \bar{\bm{B}} \bar{\bm{Z}})\\
		\text{s.t.} &\quad  \bar{Z}_{nn}=1,\quad n=1,\ldots, \bar{N}_{s}+1\\
		& \quad \bar{\bm{Z}}\succeq 0\, ,  
		\quad \bar{\bm{Z}}= \bar{\bm{z}} \bar{\bm{z}}^{\herm} 
	\end{aligned}
\end{equation}
that admits the following (convex) relaxation
\begin{equation}\label{problem-x-design_4}
	\begin{aligned}
		\max_{\bar{\bm{Z}}} & \quad \trace( \bar{\bm{B}} \bar{\bm{Z}})\\
		\text{s.t.} & \quad \bar{Z}_{nn}=1,\quad n=1,\ldots, \bar{N}_{s}+1 \, , 
		\quad \bar{\bm{Z}}\succeq 0
	\end{aligned}
\end{equation}
which can be solved by standard techniques, such as interior point methods, first-order methods on the associated dual problem, block coordinate ascent, etc. Let $\bar{\bm{Z}}^\star$ be the solution to~\eqref{problem-x-design_4}; as shown in~\cite{Goemans_2004, Zhang_Huang_2006, Waldspurger_2015}, a sub-optimum solution $\bar{\bm{s}}$ to~\eqref{problem-x-design_2} can be obtained by generating a sample vector from a complex zero-mean circularly symmetric Gaussian distribution with covariance matrix $\bm{\bar{\bm{Z}}}^\star$ and by normalizing its entries to have a unit modulus. This randomized algorithm satisfies a noticeable property: indeed, letting $\bar{f}^{\star}$ be the optimal value of the objective function in~\eqref{problem-x-design_2}, we have~\cite{Zhang_Huang_2006, So_Zhang_Ye_2007}
\begin{equation}
     \trace(\bar{\bm{B}} \bar{\bm{Z}}^{\star}) \geq \bar{f}^{\star} \geq \text{E}[\bar{\bm{s}}^{\herm} \bar{\bm{B}} \bar{\bm{s}}] \geq \frac{\pi}{4} \trace(\bar{\bm{B}} \bar{\bm{Z}}^{\star})
\end{equation}
which shows that this is a randomized $\frac{\pi}{4}$-approximation algorithm.\footnote{A (randomized) $\alpha$-approximation algorithm is an algorithm that runs in polynomial time and produces a solution whose (expected) value is at least a fraction $\alpha$ of the optimum value~\cite{book-Hochbaum}.} Clearly, multiple (independent) feasible points $\bar{\bm{s}}$ can be computed, and the one providing the largest objective function in~\eqref{problem-x-design_2} can be chosen: in this case, the $\frac{\pi}{4}$ approximation ratio will not only hold in mean but also with a probability that goes to 1 exponentially fast with the number of samples. The complexity of this approach depends on the technique used to solve Problem~\eqref{problem-x-design_4}: e.g., $\mathcal O(\bar N_s^3)$ per iteration for interior point methods and for the block coordinate ascent algorithm, or  $\mathcal O(\bar N_s^2 \ln \bar N_s)$ per iteration for the sub-gradient algorithm on the associated dual problem~\cite{Waldspurger_2015}. Additionally, there is a cost of $\mathcal O(\bar N_s^3)$ for the evaluation of the square root matrix of $\bar{\bm Z}^\star$ needed in the randomization step.

\subsection{Far-field deployment of the \glspl{ris}}\label{SEC:Far-field}
We study here in more detail the case where the radar transmitter and the forward \gls{ris} are in each other's far-field and the radar receiver and the backward \gls{ris} are in each other's far-field. In this situation, a plane wave approximation in the aforementioned radar-\gls{ris} hops can be made and, hence, the matrices in~\eqref{Gbar-def} and~\eqref{Gddot-def} simplify to $\bar{\bm{G}} =\bar{\bm{g}}_{r}\bar{\bm{g}}_{s}^{\transp}$ and $\ddot{\bm{G}}=\ddot{\bm{g}}_{r}\ddot{\bm{g}}_{s}^{\transp}$, respectively, where $\bar{\bm{g}}_{r}$ is the steering vector of the radar transmitter towards the forward \gls{ris}, $\ddot{\bm{g}}_{r}$ is the steering vector of the radar receiver towards the backward \gls{ris}, $\bar{\bm{g}}_{s}$ is the steering vector of the forward \gls{ris} towards the radar transmitter, and $\ddot{\bm{g}}_{s}$ is the steering vector of the backward \gls{ris} towards the radar receiver: all these steering vectors are tied to the geometry of the corresponding array and have unit modulus entries. Since  the matrix 
\begin{equation}
	\bar{\bm{Q}}=\underbrace{\bar{\bm{g}}_{r}}_{\bar{\bm{a}}}\underbrace{\bar{\bm{g}}_{s}^{\transp}\diag(\bar{\bm{v}}_{s})\bar{\gamma}_{s}}_{\bar{\bm{b}}^{\transp}}
	\end{equation}
has rank one, upon exploiting~\eqref{phase-design-rank-1} and~\eqref{phase-design-rank-1-sol}, we have that 
\begin{equation}\label{opt-phi-farfield}
	\max_{\bar{\bm{x}}}\|\bar{\bm{e}}(\bar{\bm{x}})\|^2=\bar{N}_{r}|\bar{\gamma}_{r}|^2 +\bar{N}_{r}\bar{N}_{s}^{2}|\bar{\gamma}_{s}|^2+2\bar{N}_{s}\big|\bar{\gamma}_{r}^*\bar{\gamma}_{s}\big|\big|\bar{\bm{v}}_{r}^{\herm}\bar{\bm{g}}_{r}\big|
\end{equation} 
that is achieved when
\begin{equation}\label{phi-far-field}
	\bar{\varphi}_{n}=-\angle\bigl(\bar{g}_{s,n}\bar{v}_{s,n}\bar{\gamma}_{s}\bigr)- \angle \bigl(\bar{\gamma}_{r}^{*} \bar{\bm{v}}_{r}^{\herm}\bar{\bm{g}}_{r}\bigr),\quad n=1,\ldots,\bar{N}_{s}.
	\end{equation}
Similarly, we have that
\begin{equation} \label{opt-barphi-farfield}   
\max_{\ddot{\bm{x}}}\|\ddot{\bm{e}}(\ddot{\bm{x}})\|^2=
\ddot{N}_{r}|\ddot{\gamma}_{r}|^2 +\ddot{N}_{r}\ddot{N}_{s}^{2}|\ddot{\gamma}_{s}|^2+2\ddot{N}_{s} \big|\ddot{\gamma}_{r}^*\ddot{\gamma}_{s}\big| \big|\ddot{\bm{v}}_{r}^{\herm}\ddot{\bm{g}}_{r}\big|
\end{equation}
that is achieved when
\begin{equation}\label{barphi-far-field}
	\ddot{\varphi}_{n}=-\angle\bigl(  \ddot{g}_{s,n} \ddot{v}_{s,n}\ddot{\gamma}_{s} \bigr)-\angle \bigl( \ddot{\gamma}_{r}^* \ddot{\bm{v}}_{r}^{\herm}\ddot{\bm{g}}_{r}\bigr),\quad n=1,\ldots,\ddot{N}_{s}.
\end{equation}

To get some insights into the achievable performance, we now evaluate the optimal \gls{snr} under a \gls{los} radar configuration (i.e., $\bar{\gamma}_{r}\ddot{\gamma}_{r}\neq0$); from~\eqref{Kronecker-structure}, \eqref{output-snr},  \eqref{opt-phi-farfield}, and~\eqref{opt-barphi-farfield}, we obtain
\begin{align}
	\text{SNR}&=\underbrace{\frac{ \bar{N}_{r}\ddot{N}_{r} | \bar{\gamma}_{r}\ddot{\gamma}_{r}|^2 \sigma_{\alpha}^2}{\sigma^2_{w}}}_{\text{SNR}_{o}}  \notag \\ 
	& \quad \times \underbrace{\Bigg(1 +\bar{N}_{s}^{2}\frac{|\bar{\gamma}_{s}|^2}{|\bar{\gamma}_{r}|^2}+2\bar{N}_{s}\frac{|\bar{\gamma}_{s}|}{|\bar{\gamma}_{r}|} \frac{\big|\bar{\bm{v}}_{r}^{\herm}\bar{\bm{g}}_{r}\big|}{\bar{N}_{r}}\Bigg)}_{\bar{\Gamma}} \notag \\
		&\quad \times \underbrace{\Bigg(1+\ddot{N}_{s}^{2}\frac{|\ddot{\gamma}_{s}|^2}{|\ddot{\gamma}_{r}|^2}+2\ddot{N}_{s}\frac{|\ddot{\gamma}_{s}|}{|\ddot{\gamma}_{r}|} \frac{\big|\ddot{\bm{v}}_{r}^{\herm}\ddot{\bm{g}}_{r}\big|}{\ddot{N}_{r}}\Bigg)}_{\ddot{\Gamma}}	
	\end{align}
where $\text{SNR}_{o}$ is the \gls{snr} when the radar operates alone, while $\bar{\Gamma}>1$ and $\ddot{\Gamma}> 1$ are the  gain granted by the \gls{ris}-aided illumination and observation, respectively. Under the model in~\eqref{gamma-def}, we have 
\begin{subequations}\label{gain_tx_rx}
\begin{align}
	\bar{\Gamma}&\approx 
	1 +  \frac{\bar{\mathcal{G}}(\bar{\rho})\bar{L}_{r}}{\bar{\mathcal{G}}(\bar{\theta}_{r}) \bar{L}_{s}}  \frac{\bar{N}_{s}^{2} \bar{\zeta}(\bar{\omega},\bar{\theta}_{s})  }{4 \pi\bar{\delta}^2} \notag 
	\\ & \quad  +2  \left(\frac{\bar{\mathcal{G}}(\bar{\rho})\bar{L}_{r}}{\bar{\mathcal{G}}(\bar{\theta}_{r}) \bar{L}_{s}}  \frac{\bar{N}_{s}^{2} \bar{\zeta}(\bar{\omega},\bar{\theta}_{s})  }{4 \pi\bar{\delta}^2} \right)^{1/2}    \frac{|\bar{\bm{v}}_{r}^{\herm}\bar{\bm{g}}_{r}|}{\bar{N}_{r}}\label{gain_tx}\\
	\ddot{\Gamma}&\approx 
	1 + \frac{\ddot{\mathcal{G}}(\ddot{\rho})\ddot{L}_{r}}{\ddot{\mathcal{G}}(\ddot{\theta}_{r}) \ddot{L}_{s}}  \frac{\ddot{N}_{s}^{2}\ddot{\zeta}(\ddot{\theta}_{s},\ddot{\omega})}{4 \pi\ddot{\delta}^2} \notag \\ 
	& \quad +2  \left(\frac{\ddot{\mathcal{G}}(\ddot{\rho})\ddot{L}_{r}}{\ddot{\mathcal{G}}(\ddot{\theta}_{r}) \ddot{L}_{s}}  \frac{\ddot{N}_{s}^{2}\ddot{\zeta}(\ddot{\theta}_{s},\ddot{\omega})}{4 \pi\ddot{\delta}^2} \right)^{1/2}    \frac{|\ddot{\bm{v}}_{r}^{\herm}\ddot{\bm{g}}_{r}|}{\ddot{N}_{r}}\label{gain_rx}
\end{align}
\end{subequations}
where the above approximations follow from the fact that $\bar{d}_{r}\approx \bar{d}_{s}$ and $\ddot{d}_{r}\approx \ddot{d}_{s}$, respectively. Observe now that,  we must necessarily have  $\bar{\delta}\lambda >2\bar{D}_{s}^{2}$ and $\ddot{\delta}\lambda >2\ddot{D}_{s}^{2}$ in order to ensure far-field operation~\cite{book-Stutzman03}; consequently, assuming that square \glspl{ris} are employed, so that $\bar{N}_{s}\bar{A}=\bar{D}_{s}^{2}$ and $\ddot{N}_{s}\ddot{A}=\ddot{D}_{s}^{2}$, under the model in~\eqref{bistatic_rcs} we have
\begin{subequations}\label{cond_tx_rx}
\begin{align}\label{cond_tx}
\frac{\bar{N}_{s}^{2}\bar{\zeta}(\bar{\omega},\bar{\theta}_{s})}{4 \pi\bar{\delta}^2 }&\leq \left(\frac{\bar{N}_{s}\bar{A}}{\bar{\delta}\lambda}\right)^2\leq \frac{1}{4}\\
\frac{\ddot{N}_{s}^{2}\ddot{\zeta}(\ddot{\theta}_{s},\ddot{\omega})}{4 \pi\ddot{\delta}^2 }&\leq \left(\frac{\ddot{N}_{s}\ddot{A}}{\ddot{\delta}\lambda}\right)^2\leq \frac{1}{4}.\label{cond_rx}
\end{align}
\end{subequations} 
From~\eqref{gain_tx_rx} and~\eqref{cond_tx_rx} we conclude that, when $\bar{\gamma}_{r}\ddot{\gamma}_{r}\neq0$, the additional paths enabled by a far-field \gls{ris} deployment only provide a marginal \gls{snr} gain: the reason is that the reflecting area offered by the \gls{ris} cannot be sufficiently large to compensate for the signal attenuation experienced in the radar-\gls{ris} hop. As also confirmed by the analysis in Sec.~\ref{Sec:analysis}, the forward (backward) \gls{ris} should better be placed in the near-field of  the radar transmit (receive) array,  as close as possible, to significantly improve the system performance. Similar conclusions have been reached in communication-oriented applications, when the effects of the double-fading attenuation and of the bistatic \gls{rcs} of the reflecting elements are accounted for in the \gls{ris}-aided link~\cite{Najafi-2021,ActiveRIS2021}.

We hasten to underline that, when the  \gls{ris} is in the far-field of both the signal source and the destination, many existing works have bypassed the fact that the indirect link can be much weaker than the direct one by assuming that the latter is obstructed; indeed, such a far-field deployment may remain a competitive option if the indirect link allows to reach a user that would otherwise remain disconnected (in communication applications) or a spot  that would otherwise remain blind (in radar applications), in spite of the double-fading attenuation.

\subsection{Mono-static radar with a bi-directional \gls{ris}}
We study here in more detail a monostatic radar  wherein the same \gls{ris} helps both the transmitter and the receiver: in this case, we have $\bar{N}_{s}=\ddot{N}_{s}=N_{s}$, $(\bar{d}_{s},\bar{\theta}_{s})=(\ddot{d}_{s},\ddot{\theta}_{s})$, $\bar{\bm{v}}_{s}=\ddot{\bm{v}}_{s}$, $\bar{\gamma}_{s}\ddot{\gamma}_{s}\neq0$, and $\bar{\zeta}(\cdot, \cdot)=\ddot{\zeta}(\cdot, \cdot)$. The disjoint design discussed in Section~\ref{Sec:disjoint-phase-design} can directly be  employed if the phase shift introduced by each reflecting element can be reprogrammed in between the arrival of the forward (from the radar transmitter) and of the backward  (from the target) incident wave. To simplify the hardware, we also discuss next the relevant case where the additional constraint $\bar{\bm{x}}=\ddot{\bm{x}}=\bm{x}$ is enforced to~\eqref{problem-coherent-2}; accordingly, the design of the phase shifts is recast as
\begin{equation}\label{problem-eqPHI-design}
	\max_{\bm{x}}\; \|\bm{e}(\bm{x},\bm{x})\|^2,\quad
	\text{s.t. } |{x}_{n}|=1,\;n=1,\ldots,N_{s}
\end{equation}
Notice that forcing $\bar{\bm{x}}=\ddot{\bm{x}}$ is  optimal if $\bar{\bm{e}}(\cdot)=\ddot{\bm{e}}(\cdot)$: this occurs when the \gls{ris} is reciprocal \emph{and} the radar employs the same reciprocal array for both transmission and reception; instead, it is inevitably sub-optimal if the system presents an asymmetry between the transmit and the receive side. In this latter case, we  compute a sub-optimal solution to~\eqref{problem-eqPHI-design} by iteratively optimizing one phase shift at a time. To proceed,  the objective function is expanded as 
\begin{align}
	\| \bm{e}(\bm{\varphi},\bm{\varphi})\|^2&= \left\| \bar{\bm{t}}_{n}+\bar{\bm{q}}_{n} x_{n}\right\|^2 \left\| \ddot{\bm{t}}_{n}+\ddot{\bm{q}}_{n} x_{n}\right\|^2\notag\\
	&=	\bigl(\| \bar{\bm{t}}_{n}\|^2+\| \bar{\bm{q}}_{n}\|^2\bigr)\bigl(\| \ddot{\bm{t}}_{n}\|^2+\| \ddot{\bm{q}}_{n}\|^2\bigr)\notag\\
	& \quad + 2\bigl(\| \bar{\bm{t}}_{n}\|^2+\| \bar{\bm{q}}_{n}\|^2\bigr)\Re\left\{ \ddot{\bm{t}}_{n}^{\herm}\ddot{\bm{q}}_{n} x_{n}\right\}	\notag\\
	& \quad + 2\bigl(\| \ddot{\bm{t}}_{n}\|^2+\| \ddot{\bm{q}}_{n}\|^2\bigr) \Re\left\{ \bar{\bm{t}}_{n}^{\herm}\bar{\bm{q}}_{n}x_{n}\right\}\notag\\
	&\quad + 4\Re\left\{ \ddot{\bm{t}}_{n}^{\herm}\ddot{\bm{q}}_{n} \bar{\bm{t}}_{n}^{\herm}\bar{\bm{q}}_{n} x_{n}^{2}\right\}\label{eq:obj-exp-eq}
\end{align}
where $\bar{\bm{q}}_n$ is the $n$-th column of the matrix $\bar{\bm{G}}\diag\left(\bar{\bm{v}}_{s}\right)\bar{\gamma}_{s}$, and $\bar{\bm{t}}_{n}=\bar{\gamma}_{r}\bar{\bm{v}}_{r}+\sum_{j=1,\,j\neq n}^{N_{s}}\bar{\bm{q}}_{j}x_{j}$ at the transmit side, and $\ddot{\bm{q}}_n$ is the $n$-th column of the matrix $\ddot{\bm{G}}\diag\left(\ddot{\bm{v}}_{s}\right)\ddot{\gamma}_{s}$ and $\ddot{\bm{t}}_{n}=\ddot{\gamma}_{r}\ddot{\bm{v}}_{r}+\sum_{j=1,\,j\neq n}^{N_{s}}\ddot{\bm{q}}_{j}x_{j}$ at the receive side. Neglecting the irrelevant terms in~\eqref{eq:obj-exp-eq}, the problem to be solved becomes 
\begin{equation}\label{eq:obj-exp-eq1}
	\max_{\varphi_n}\; \Re\bigl\{A_{n} e^{\i\varphi_n}+B_n e^{\i2\varphi_n} \bigr\}
\end{equation}
where $x_{n}=e^{\i\varphi_n}$,	$A_n= 2\bigl(\| \bar{\bm{t}}_{n}\|^2+\| \bar{\bm{q}}_{n}\|^2\bigr)\ddot{\bm{t}}_{n}^{\herm}\ddot{\bm{q}}_{n}+2\bigl(\| \ddot{\bm{t}}_{n}\|^2+\| \ddot{\bm{q}}_{n}\|^2\bigr)\bar{\bm{t}}_{n}^{\herm}\bar{\bm{q}}_{n}$ and $
B_n=4\ddot{\bm{t}}_{n}^{\herm}\ddot{\bm{q}}_{n} \bar{\bm{t}}_{n}^{\herm}\bar{\bm{q}}_{n}$. The overall procedure is summarized in Algorithm~\ref{alg_eqphi_design}: since the value of the objective function is bounded above and monotonically increased at each iteration, convergence is ensured. Finally, notice that the solution to~\eqref{eq:obj-exp-eq1} can be found among the zeros of the  derivative of the objective function, i.e., $
	|A_{n}|\sin\left(\varphi_{n}+\angle A_{n}\right) +2|B_{n}|\sin\left(2\varphi_{n}+\angle B_{n}\right)$,
which can be obtained numerically. 
\begin{algorithm}[t]
	\caption{Alternate maximization for Problem~\eqref{problem-eqPHI-design}}
	\begin{algorithmic}[1]\label{alg_eqphi_design}
		\STATE Choose $\epsilon>0$, $K _\text{max}>0$, and $\bm{x}\in \mathbb{C}^{N_{s}}$: $|x_{n}|=1$, $\forall n$
		\STATE $k=0$ and $f_{0}=0$ 
		\REPEAT
		\FOR{$n=1,\ldots,N_{s}$}
		\STATE 	$\displaystyle \angle x_n=\arg\max_{\varphi} \Re\bigl\{A_{n} e^{\i\varphi}+B_n e^{\i2\varphi} \bigr\}$
		\ENDFOR
		\STATE $k = k+1$
		\STATE $f_{k}=\|\bm{e}(\bm{x},\bm{x})\|^2$
		\UNTIL{$f_{k}-f_{k-1}<\epsilon f_{k}$ or $k=K_\text{max}$}
	\end{algorithmic}
\end{algorithm}

\subsection{More insights on the radar and \gls{ris} interplay}
An \gls{mimo} radar emitting orthogonal waveforms simultaneously illuminates a large angular sector, whose width is tied to the beampattern of the transmit antennas; when operating alone, this radar can be focused on a desired point in the illuminated region by only relying on receive signal processing, and multiple points can be  simultaneously inspected by changing the range gate and the receive filter.

An \gls{ris}-aided \gls{mimo} radar benefits from additional degrees of freedom, namely, the choice of the phase shifts of the reflecting elements and a modified field of view. Since an \gls{ris}  must be programmed  before the transmission of the radar waveforms, the system engineer must decide in advance how it should operate.  The  design proposed in~\eqref{problem-coherent-2} chooses the phase shifts of both the forward and the backward \gls{ris} in order to enhance the \gls{snr} at only one desired location. For example, the location of interest can be  the one where an alert was previously found (if the radar is operating in an alert/confirm mode~\cite{Grossi-two-step-2017})  or the one where a previously-detected target is expected to be found  (if the radar is operating in a tracking mode) or a spot not accessible by the radar when operating alone~\cite{aubry2021reconfigurable}. Even if the \glspl{ris} are optimized to look at only one location, the radar can still simultaneously inspect other spots.

Needless to say, different design criteria could be also considered to operate  the \gls{ris}-aided \gls{mimo} radar. For example, the forward and backward \glspl{ris} could be steered towards multiple, possibly different locations, which may be of interest in the presence of multiple targets. 

\section{Performance analysis}\label{Sec:analysis}
We present here some examples to assess the performance achievable by an \gls{ris}-aided \gls{mimo} radar. The analysis is carried out under the model in Sec.~\ref{Sec:channel}, with $\bar{L}_{r}=\bar{L}_{s}=\ddot{L}_{r}=\ddot{L}_{s}$ in~\eqref{gamma-def}. The \gls{ris} design is undertaken by resorting to Algorithms~\ref{alg_x_design} and~\ref{alg_eqphi_design}, which are implemented with $K_{\max}=200$, $\epsilon=10^{-5}$, and $100$ random initialization points.\footnote{In the following examples, the alternate maximization and the semi-definite relaxation with randomization discussed in Section~\ref{Sec:disjoint-phase-design} have provided similar values of the objective function in Problem~\eqref{problem-x-design_2}.}

\subsection{System geometry}\label{Sec:numerical-analysis-geometry}
\begin{figure}[t]	
	\centering	
	\centerline{\includegraphics[width=0.8\columnwidth]{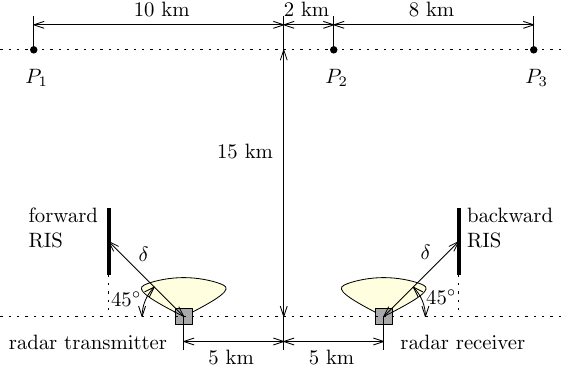}}	
	\vspace{-0.2cm}\caption{Considered system geometry. The radar transmitter and receiver and the forward and backward RISs are equipped with planar arrays; their local $x$ and $y$ axes lay on the same plane, which also contains the array centers and the prospective target. The target is either fixed at $P_{2}$ or moved along the line segment connecting $P_{1}$ to $P_{3}$ (from left to right).} \label{simulated_geometry}	
\end{figure}
We consider the geometry in Fig.~\ref{simulated_geometry}, which features a bistatic radar operating at $3$~GHz, aided by a forward and/or backward \gls{ris}.\footnote{The S-band ($2\div4$ GHz)  is used for example by airport surveillance radars, weather radars, and surface ship radars, which can cover up to $100$~km.} The radar transmitter and receiver employ a linear array with a $\lambda/2$ element spacing along the local $y$-axis, $\bar{N}_{r}=3$, and (unless otherwise stated) $\ddot{N}_{r}=8$; each array element has a rectangular shape of size $\lambda/2$ and $\lambda$ along the local $y$ and $z$ axes, respectively, and a power beampattern with a $3$-dB width of $120^\circ$ in azimuth and of $60^\circ$ in elevation. The \glspl{ris} (if present) have a square shape and are composed of $\bar{N}_{s}=\ddot{N}_{s}=N_{s}$ adjacent square elements with an area $\bar{A}=\ddot{A}=\lambda^2/4$ and a cosine exponent $q=1$ in~\eqref{bistatic_rcs}: this may for example correspond to a reflectarray-based \gls{ris} implementation~\cite{Geoffrey-Ye-2020}. 
The reference element of each array is at the center of the array, with $\bar{\delta}=\ddot{\delta}=\delta$, while the target has an effective size of $\bar{D}_{t}=\ddot{D}_{t}=10\lambda$. Unless otherwise stated, we report next the performance when the target is at $P_{2}$ and the \glspl{ris} are designed to maximize the \gls{snr} at this same location.

\subsection{Impact of the radar-\gls{ris} distance}
\begin{figure}[t]	
	\centering	
	\centerline{\includegraphics[width=\columnwidth]{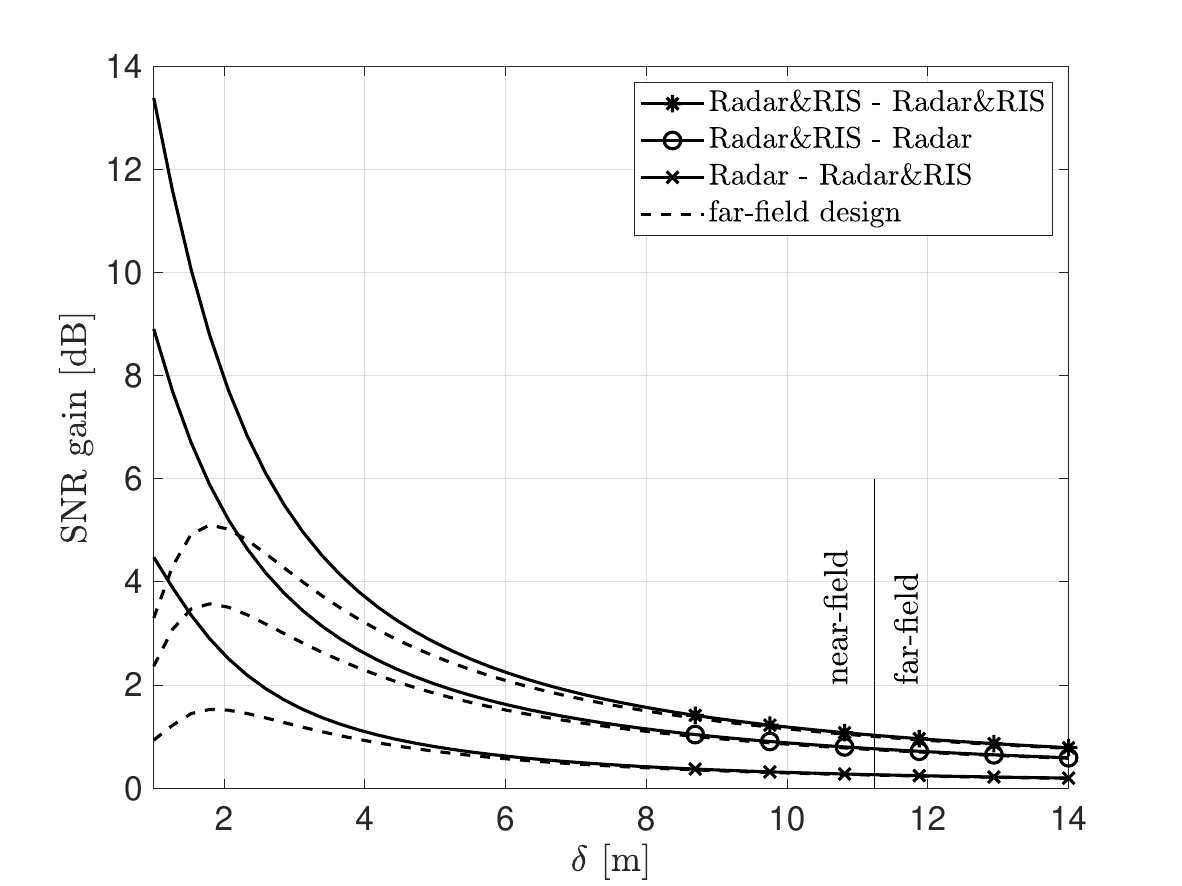}}	
	\vspace{-0.2cm}\caption{\gls{snr} gain (as compared to case where the radar operates alone) obtained by using a forward and/or a backward \gls{ris} versus $\delta$, when $N_{s}=225$. The system geometry in Fig.~\ref{simulated_geometry} is considered with the target located at $P_{2}$.} \label{fig:LOSbi_distance_1}	
\end{figure}
\begin{figure}[t]	
	\centering	
	\centerline{\includegraphics[width=\columnwidth]{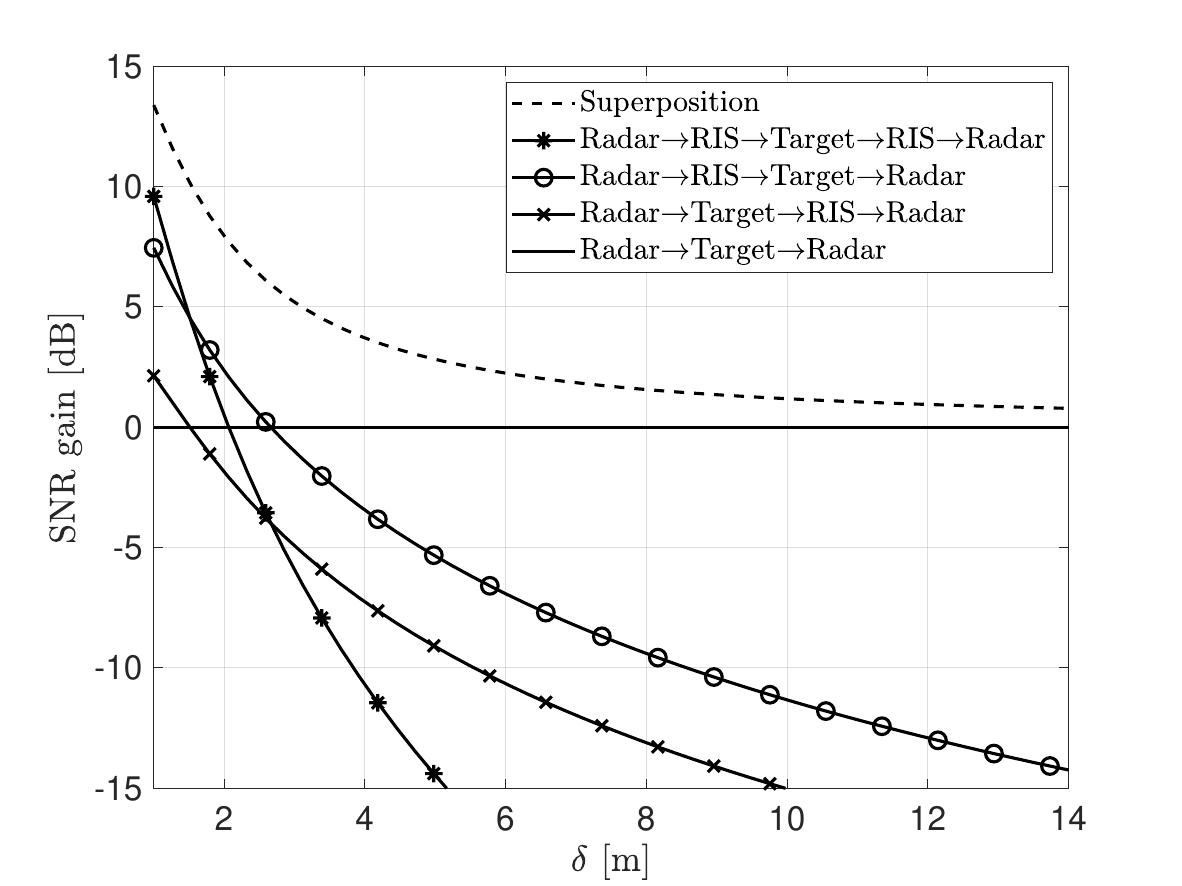}}	
	\vspace{-0.2cm}\caption{\gls{snr} gain (as compared to case where the radar operates alone) of the  the four echoes in~\eqref{Kronecker-structure} and of their superposition versus $\delta$, when both \gls{ris} are employed (``Radar\&RIS - Radar\&RIS'' case) and $N_{s}=225$. The system geometry in Fig.~\ref{simulated_geometry} is considered  with the target located at $P_{2}$.} \label{fig:LOSbi_distance_2}	
\end{figure}
\begin{figure}[t]	
	\centering	
	\centerline{\includegraphics[width=\columnwidth]{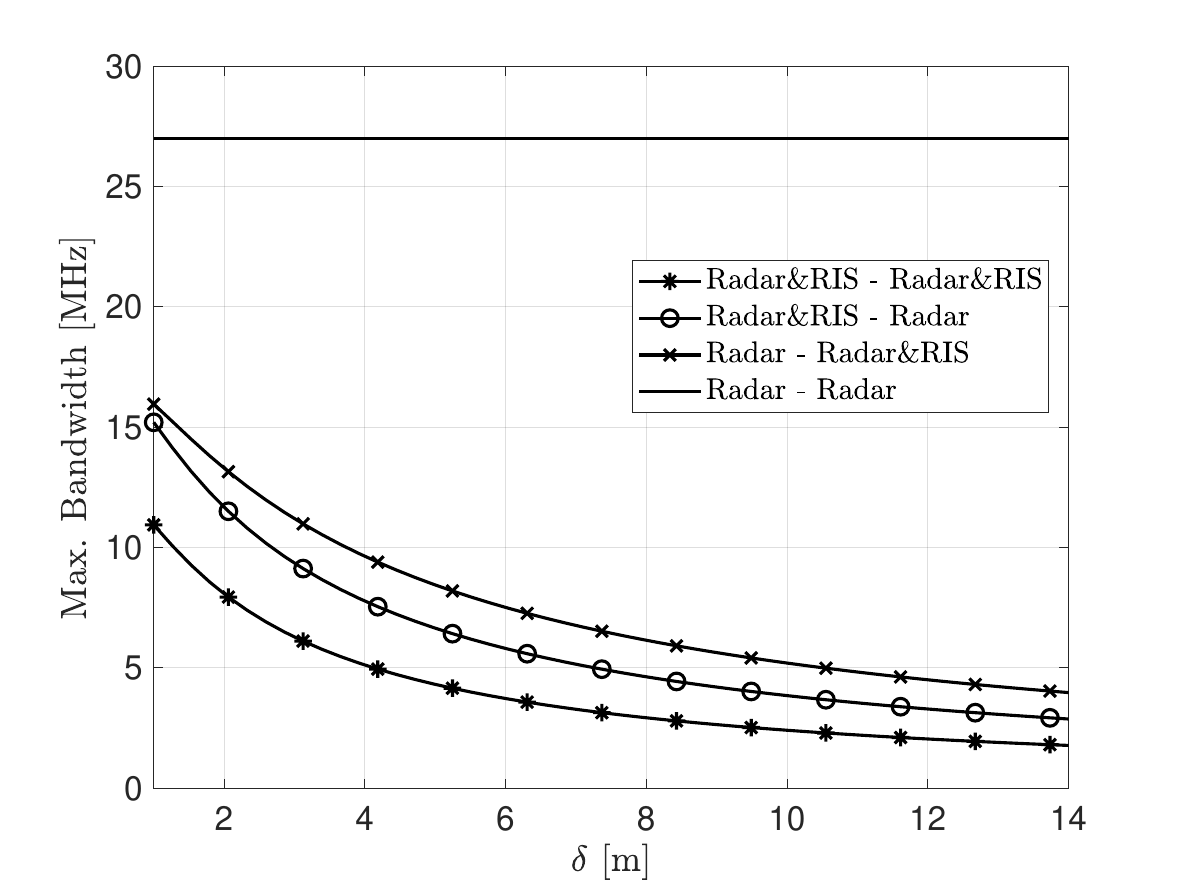}}	
	\vspace{-0.2cm}\caption{Maximum bandwidth ensuring that the delays of all target echoes reaching the receiver are not resolvable versus $\delta$. The system geometry in Fig.~\ref{simulated_geometry} is considered  with the target located at $P_{2}$.} \label{fig:LOSbi_distance_3}	
\end{figure}
We first study the system behavior when $\delta$ is varied and $N_{s}=225$. Fig.~\ref{fig:LOSbi_distance_1} reports  the \gls{snr} gain obtained by using a forward and a backward \gls{ris}, a forward \gls{ris} only, and a backward \gls{ris} only, as compared to the case where the radar operates alone (i.e., only the ``Radar$\rightarrow$Target$\rightarrow$Radar'' echo is present): the configurations considered here correspond to the ``Radar\&RIS - Radar\&RIS'', ``Radar\&RIS - Radar", and ``Radar - Radar\&RIS" cases\footnote{The label ``Radar - Radar\&RIS'' indicates that the target is illuminated by the radar and observed by both the radar and the RIS; the meaning of ``Radar\&RIS - Radar\&RIS'' and  ``Radar\&RIS - Radar" is similarly understood.} reported in Table~\ref{table:configurations}. For comparison, we include the performance obtained with the far-field design in~\eqref{phi-far-field} and~\eqref{barphi-far-field}. It is seen that a larger \gls{snr} gain is obtained when the radar and \gls{ris} get closer, and, in this regime, the far-field design turns out to be detrimental. The \gls{snr} gain becomes instead negligible when the radar and \gls{ris} arrays are located in each other's far-field at both the transmit and the receive side, which approximately occurs when $\delta\geq 11.5$~m in this example (as indicated by the vertical dotted line). Simultaneously using a forward and a backward \gls{ris} almost doubles the gain as compared to using a single reflecting array. If only a single \gls{ris} has to be employed (e.g., due to cost constraints), then, for the considered scenario, it should be that at the transmit side: this is a consequence of the fact that the reflecting elements of the forward \gls{ris} offer a larger bistatic \gls{rcs} towards $P_{2}$ than those of the forward counterpart.  To get further insights,  in Fig.~\ref{fig:LOSbi_distance_2} we consider the case where both \gls{ris}s are simultaneously employed and report the \gls{snr} gain of the four echoes in~\eqref{Kronecker-structure} and of their superposition. The ``Radar$\rightarrow$RIS$\rightarrow$Target$\rightarrow$Radar'' echo is stronger than ``Radar$\rightarrow$Target$\rightarrow$RIS$\rightarrow$Radar'' echo, in keeping with the fact that the elements of the forward \gls{ris} offer a larger bistatic \gls{rcs} towards $P_{2}$; also, the ``Radar$\rightarrow$RIS$\rightarrow$Target$\rightarrow$RIS$\rightarrow$Radar'' echo is dominant when the distance between the radar and \gls{ris} arrays is very small, but it more rapidly fades out  when $\delta$ is increased due to the more severe path-loss. Finally, Fig.~\ref{fig:LOSbi_distance_3} reports the maximum bandwidth compatible with the required narrow-band assumption; this value have been computed as $0.1/(\tau_{\max}-\tau_{\min})$, where $\tau_{\max}$ and $\tau_{\min}$ are the maximum and the minimum propagation delays from the transmit to the receive elements. Not surprisingly, the maximum bandwidth decreases as $\delta$ and/or the number of hops are increased. 

\subsection{Impact of the \gls{ris} size}
\begin{figure}[t]	
	\centering	
	\centerline{\includegraphics[width=\columnwidth]{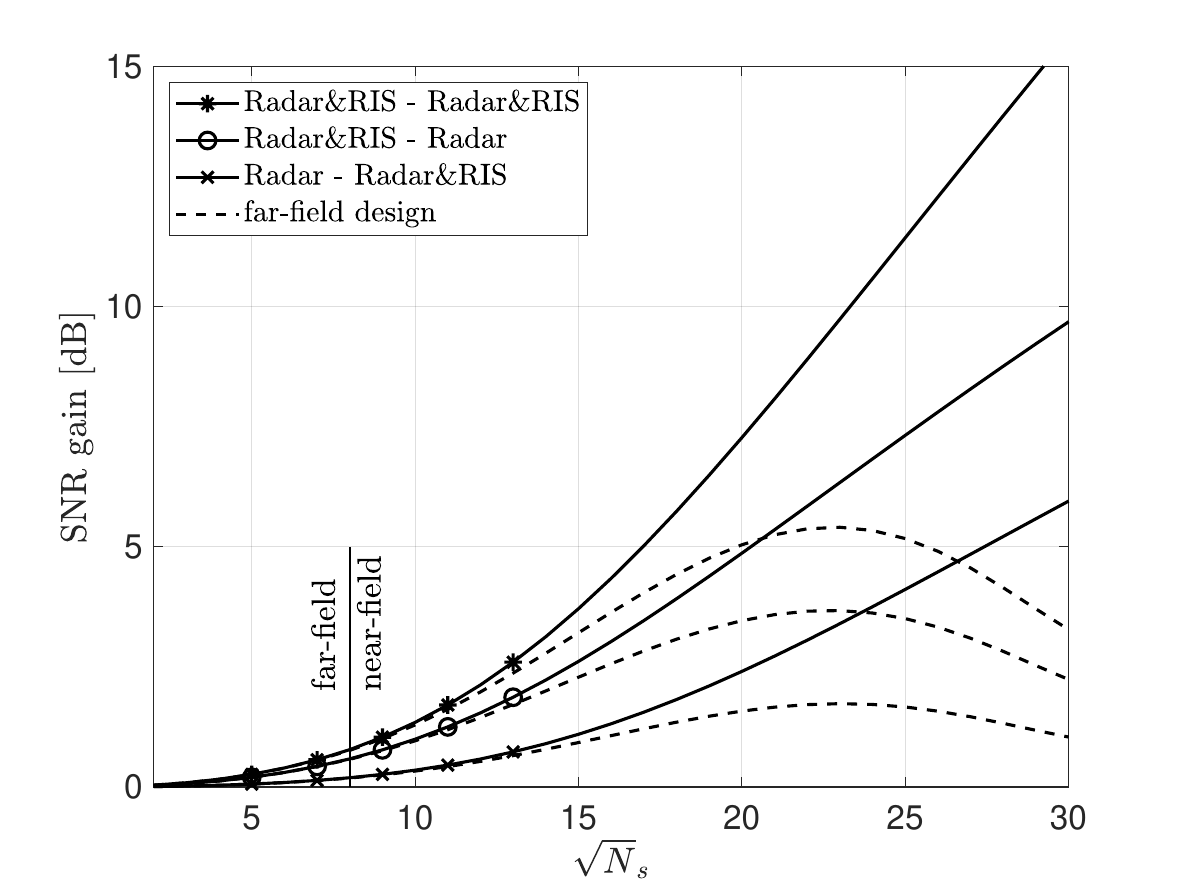}}	
	\vspace{-0.2cm}\caption{\gls{snr} gain (as compared to case where the radar operates alone)  obtained by using a forward and/or a backward \gls{ris} versus $\sqrt{N_{s}}$, when $\delta=4$~m. The system geometry in Fig.~\ref{simulated_geometry} is considered  with the target located at $P_{2}$.} \label{fig:LOSbi_size_1}	
\end{figure}
\begin{figure}[t]	
	\centering	
	\centerline{\includegraphics[width=\columnwidth]{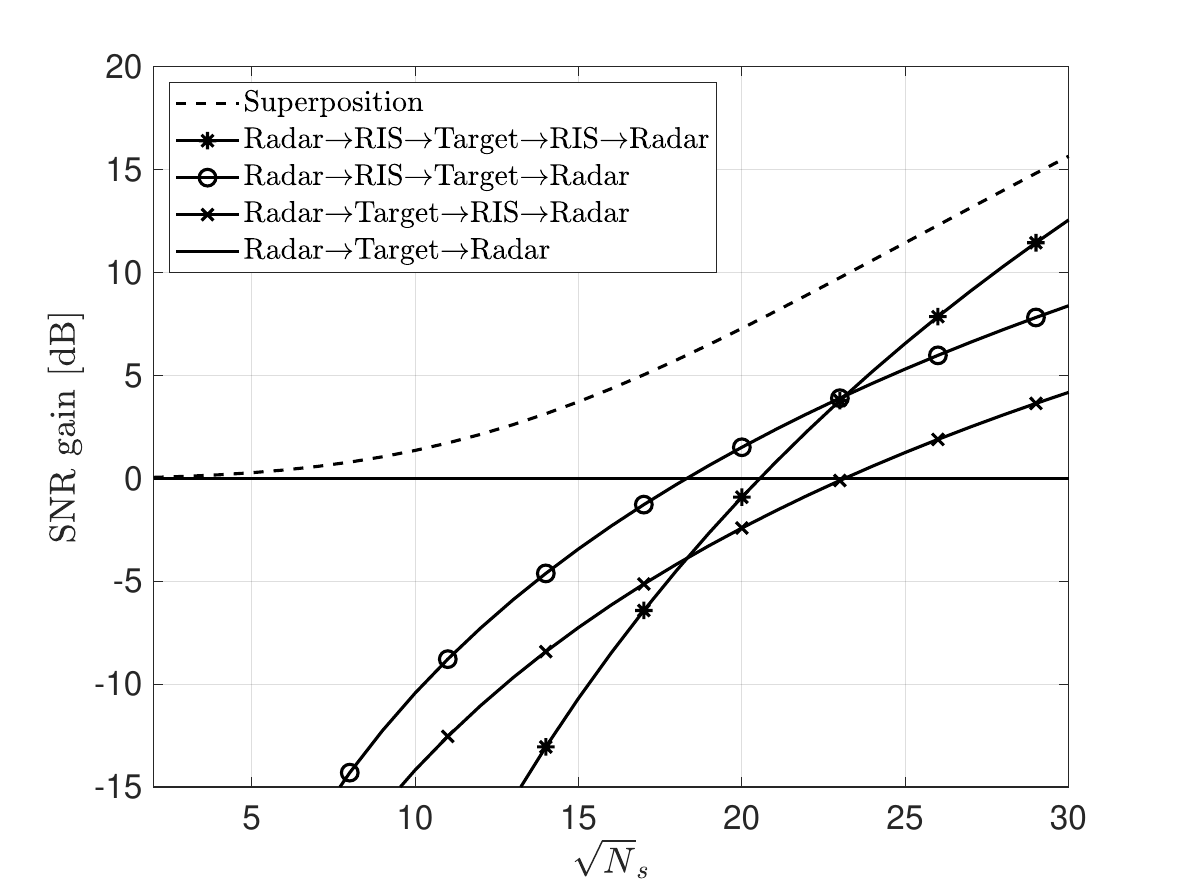}}	
	\vspace{-0.2cm}\caption{\gls{snr} gain (as compared to case where the radar operates alone) of the four echoes in~\eqref{Kronecker-structure} and of their superposition versus $\sqrt{N}_{s}$, when $\delta=4$~m and both \gls{ris} are employed (``Radar\&RIS - Radar\&RIS'' case). The system geometry in Fig.~\ref{simulated_geometry} is considered  with the target located at $P_{2}$.} \label{fig:LOSbi_size_2}	
\end{figure}

Next, we study the system behavior when the number of reflecting elements along each dimension, i.e.,  $\sqrt{N_{s}}$, is varied. Fig.~\ref{fig:LOSbi_size_1} reports the \gls{snr} gain granted by the use of a forward and/or a backward \gls{ris}, as compared to the case where the radar operates alone; the gain increases as the reflecting surfaces get larger; however, it is important to notice that, as long as $\sqrt{N_{s}}<8$, the radar and \gls{ris} arrays remain in each other's far-field (as indicated by the vertical dotted line) and only marginal gains are obtained; when instead $\sqrt{N_{s}}$ gets larger, the advantage of using one or two \glspl{ris} becomes evident. To farther investigates this phenomenon,  in Fig.~\ref{fig:LOSbi_size_2} we consider the case where both \glspl{ris} are simultaneously employed and report the \gls{snr} gain of the four echoes in~\eqref{Kronecker-structure} and of their superposition. It is seen that the indirect echoes  become progressively stronger as the \gls{ris} size is increased and, eventually, dominate over the direct one; hence, upon fixing the distance from the radar, exploiting a reflecting surface becomes competitive only if its size is sufficiently large so as to positively balance the more severe attenuation over this path. In practice, the \gls{ris} size is limited by cost and installation constraints; also, it remains understood that complementing an existing radar with a one or more \glspl{ris} can be attractive only if the cost is inferior to that of upgrading the radar itself with a better performing transceiver. 

\subsection{Impact of the target position}\label{Sec:analysis-target-position}
\begin{figure}[t]	
	\centering	
	\centerline{\includegraphics[width=\columnwidth]{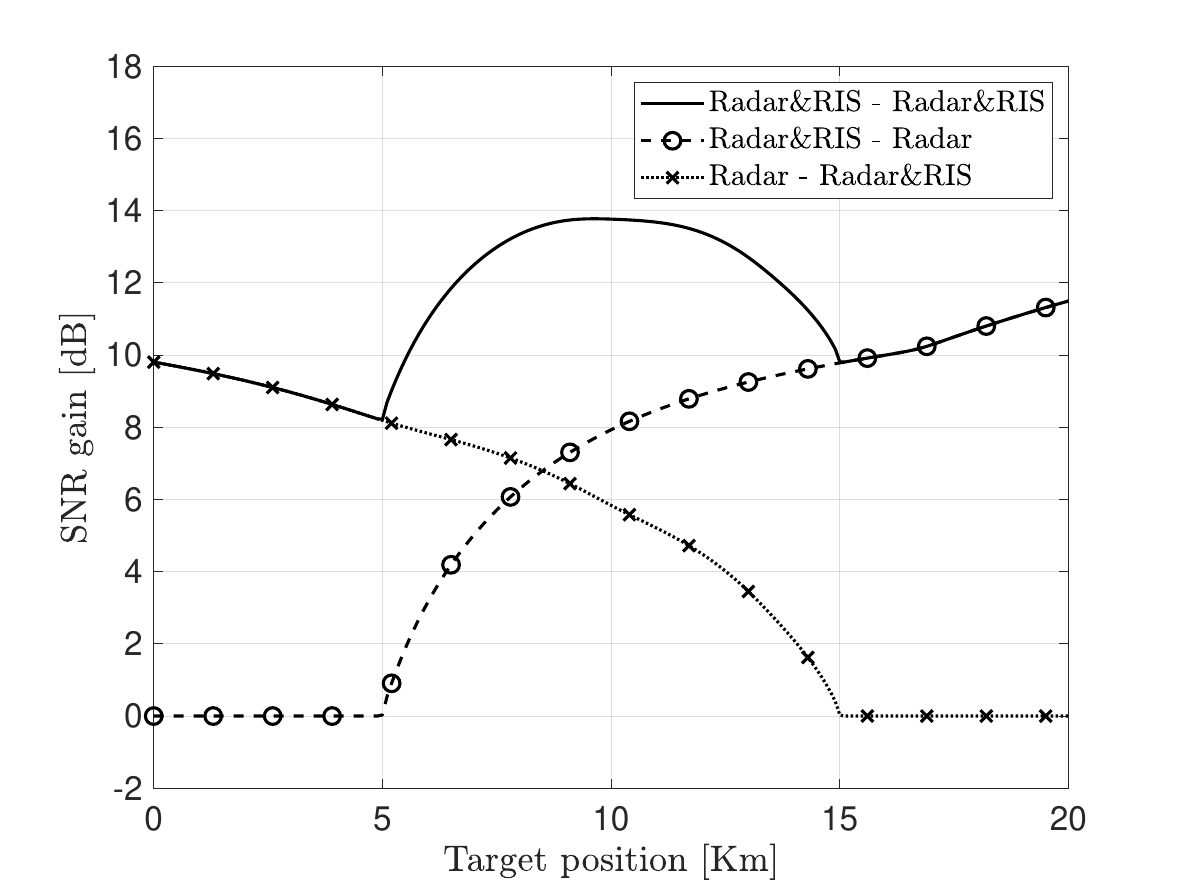}}	
	\vspace{-0.2cm}\caption{\gls{snr} gain (as compared to case where the radar operates alone)  obtained by using a forward and/or a backward \gls{ris} versus the target position along the line segment going from $P_1$ ($=0$~Km) to $P_3$ ($=20$~Km), when $N_{s}=225$, $\delta=1$~m, and the system geometry in Fig.~\ref{simulated_geometry} is considered.} \label{fig:LOSbi_posy_1}	
\end{figure}
\begin{figure*}[t]	
	\centering	
	\centerline{\includegraphics[width=0.9\textwidth]{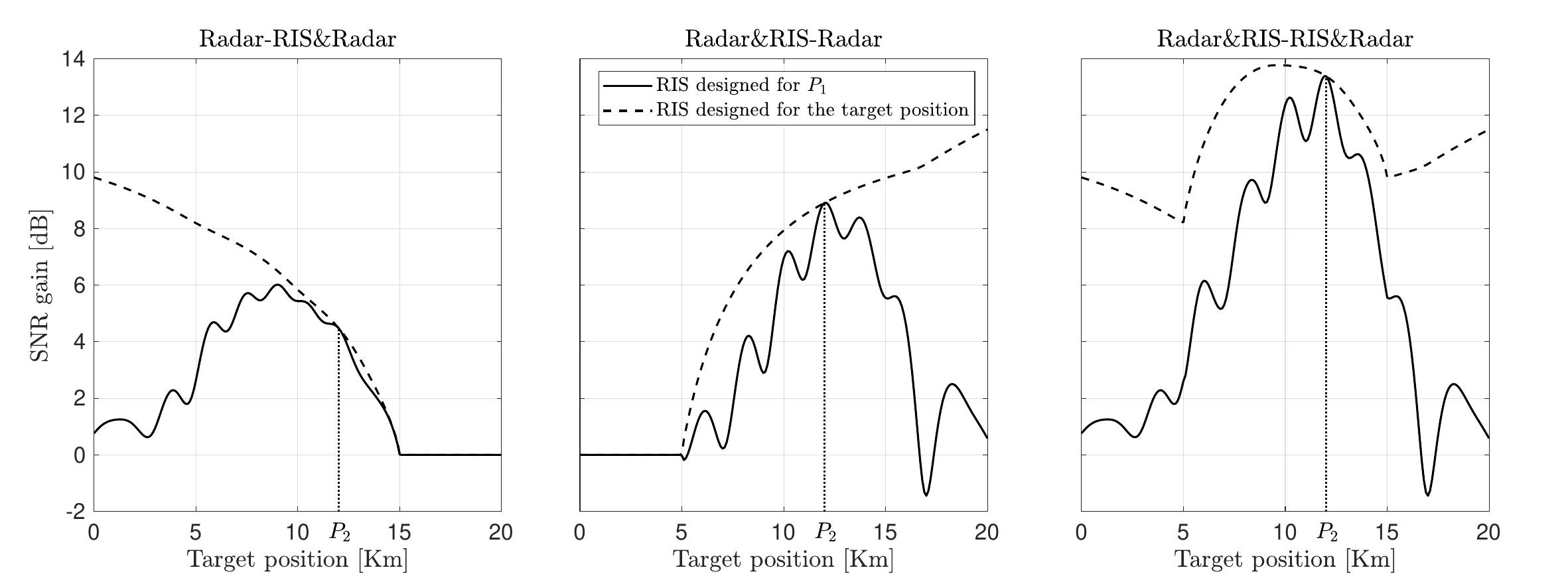}}	
	\vspace{-0.2cm}\caption{\gls{snr} gain (as compared to case where the radar operates alone)  when using a forward \gls{ris}  (left) or a backward \gls{ris}  (center) or both \gls{ris}  (right) versus the target position along the line segment going from $P_1$ ($=0$~Km) to $P_3$ ($=20$~Km), when $N_{s}=225$, $\delta=1$~m, and the system geometry in Fig.~\ref{simulated_geometry} is considered. The \glspl{ris} are designed to maximize the \gls{snr} at  $P_{2}$ ($=12$~Km) or at the actual target position. } \label{fig:LOSbi_posy_3}	
\end{figure*}

We now investigate the system performance as a function of the target position, which is moved along the line segment connecting $P_{1}$ to $P_3$ (from left to right), which has a length of $20$~Km, when $N_{s}=225$ and $\delta=1$~m. At first, we assume that the \glspl{ris} are designed to maximize the \gls{snr} at the actual target location and, in Fig.~\ref{fig:LOSbi_posy_1}, report the corresponding \gls{snr} gain. 
Notice that the forward \gls{ris} cannot illuminate the target in the interval $0\div5$~Km; accordingly, the ``Radar\&RIS--Radar'' configuration does not provide any gain here, while the ``Radar\&RIS--Radar\&RIS''  and ``Radar-Radar\&RIS'' configurations present the same gain. Similarly, the backward \gls{ris} cannot observe the target in the interval $15\div20$~Km; accordingly, the ``Radar--Radar\&RIS'' is not effective here, while the ``Radar\&RIS--Radar\&RIS''  and ``Radar--RIS\&Radar'' are equivalent. In between, the system can greatly benefit from the simultaneous use of two \glspl{ris}, in keeping with the results shown in the previous examples. The observed asymmetry in the reported gains, with respect to the middle point of the line segment, is due to the fact that $\bar{N}_{r}\neq \ddot{N}_{r}$.  Finally, Fig.~\ref{fig:LOSbi_posy_3} reports the performance when the \glspl{ris} are designed to maximize the \gls{snr} at  $P_{2}$ (marked here with a dotted vertical line); for comparison, it also reports the achievable performance when the \glspl{ris} are designed to maximize the \gls{snr} at the actual target location. Even under a mismatched design, a noticeable gain is observed across a large region around $P_{2}$, which progressively reduces by moving away; the observed ripple is due to the fact that the direct and indirect echoes may not always add constructively at locations other than $P_2$ and, indeed, a small negative gain may even occur occasionally. Overall, the results in Fig.~\ref{fig:LOSbi_posy_3} reinforce the intuition that the system engineer can opportunistically activate the available reflecting surfaces to redirect part of the radiated/received power towards/from specific spots to locally boost the detection performance, while the radar transceiver can still keep looking at other regions with acceptable loss, which appears to be a nice feature.  

\subsection{Mono-static radar configuration}
\begin{figure}[t]	
	\centering	
	\centerline{\includegraphics[width=\columnwidth]{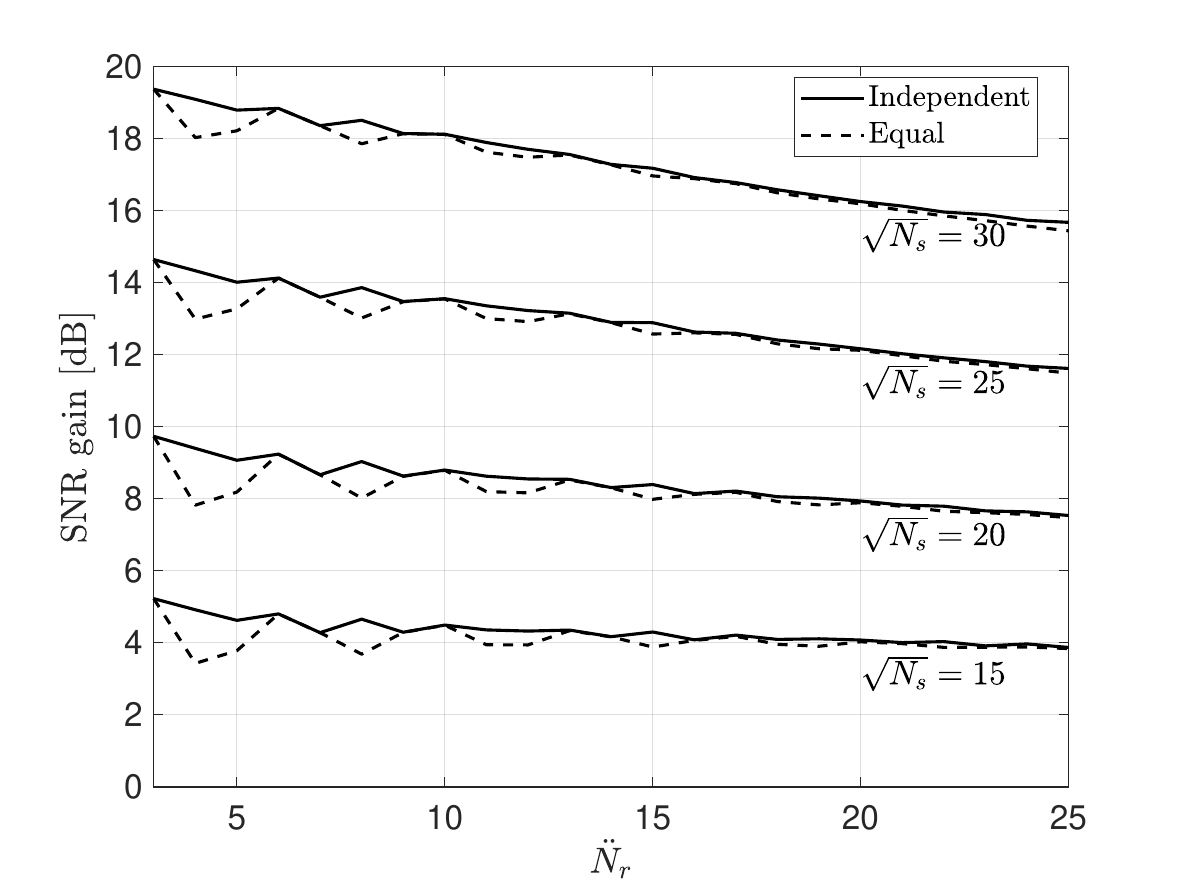}}	
	\vspace{-0.2cm}\caption{\gls{snr} gain (as compared to case where the radar operates alone) versus $\ddot{N}_{r}$,  when the forward and backward phase shifts are independently optimized or are forced to be equal, $\sqrt{N_{s}}=15,20,25,30$ and $\delta=4$~m. The system geometry in Fig.~\ref{simulated_geometry} is modified by moving the receiver to the same location of the transmitter and by using a single bi-directional \gls{ris}; the target is located at $P_{2}$} \label{fig:LOSmono_txrx}	
\end{figure}
We assume here that the receive array is co-located with the transmit array and the same \gls{ris} is employed for both forward and backward reflection; all other parameters remain defined as in Sec.~\ref{Sec:numerical-analysis-geometry}. Fig.~\ref{fig:LOSmono_txrx} reports the \gls{snr} gain obtained when the forward and backward phase shifts of each reflecting element are independently optimized or forced to be equal, as a function of the number of receive elements and for different \gls{ris} sizes, when $d=4$~m. Remarkably, enforcing equal phase values only incurs marginal losses, thus simplifying the required hardware. Also, the obtained \gls{snr} gain reduces as $\ddot{N}_{r}$ is increased; indeed, if the radar is equipped with a better receiver, the reward for using a nearby \gls{ris} will be less.

\section{Conclusions}\label{Sec:conclusions}
In this work we have started unveiling the benefits of complementing a \gls{mimo} radar with \glspl{ris} placed close to the transmitter/receiver. We have derived a  general signal model, which includes monostatic,  bistatic, \gls{los}, and  \gls{nlos} radar configurations and accounts for the presence of up to four paths from the transmitter to the prospective target to the receiver, and have recognized that the reflecting surfaces can be employed to enhance the achievable detection performance by redirecting/concentrating the radiated power towards/from a specific spot. Our analysis have confirmed a simple intuition: an \gls{ris} should be better placed as close as possible to the radar  transmitter/receiver in order to limit the additional path loss experienced by an indirect link, with a far-field deployment only providing a marginal advantage when the radar already has a direct view of the target. Also, in a monostatic configuration, the same surface can be used for both forward and backward reflection by simply maintaining the same phase shift,  which greatly simplifies its construction and control.

Future works should study the effect of the clutter on the system design and the joint design of the emitted space-time waverforms, the \glspl{ris} phase shifts, and the receive filter; also, they might consider how to point the \glspl{ris} towards multiple spots (widely or closely spaced).
Moreover, the impact of the mutual coupling among the \gls{ris} elements and among the RIS and radar arrays should be investigated; initial efforts in this direction have been done in~\cite{Gradoni-2021,Qian-2021}, where a circuit-based end-to-end model is developed for a \gls{ris}-aided communication channel; however, the \gls{ris}-aided radar channel may be substantially different, as it entails additional hops and passes through a non-cooperative object (namely, the target). Future developments might also investigate applications involving a high-resolution radar where the direct and indirect echoes  have resolvable delays or a passive/opportunistic radar exploiting an \gls{ris} to redirect the signal emitted by another (possibly non directional) source. A farther research line is studying if and how the estimation of the target parameters and the resolution of closely-spaced objects  can improve in the presence of \glspl{ris}. Finally, the use of active \glspl{ris} have been recently proposed for future 6G wireless communications~\cite{Larsson-2021,ActiveRIS2021}, and their potential in the context of radar systems remains to be investigated.

\bibliographystyle{IEEEtran} 
\bibliography{biblio}

\end{document}